\documentclass{article}

\usepackage{PRIMEarxiv}

\usepackage[utf8]{inputenc} 
\usepackage[T1]{fontenc}    
\usepackage{hyperref}       
\usepackage{url}            
\usepackage{booktabs}       
\usepackage{amsfonts}       
\usepackage{fancyhdr}       
\usepackage{graphicx}       
\graphicspath{{./}{figures_v2/}{media/}} 
\usepackage{siunitx}
\usepackage{xcolor}
\usepackage{subcaption}
\usepackage{bm}
\usepackage{mhchem}
\usepackage{amsmath}
\usepackage{makecell}
\usepackage{float}          
\usepackage{placeins}       

\title{A Liquid-Fueled Reactor Network Model for Enhanced NO\textsubscript{x} Prediction in Gas Turbine Combustors}

\author{
  Philip John, Haresh Chandrasekhar, Sourav Saha, Opeoluwa Owoyele%
  \thanks{Corresponding author: \texttt{oowoyele@lsu.edu}} \\[3pt]
  Department of Mechanical and Industrial Engineering \\
  Louisiana State University, Baton Rouge, LA 70810, USA
}

\begin{document}
\maketitle

\begin{abstract}
This study introduces a liquid-fueled reactor network (LFRN) framework for reduced-order modeling of gas turbine combustors. The proposed LFRN extends conventional gaseous-fueled reactor network methods by incorporating specialized reactors that account for spray breakup, droplet heating, and evaporation, thereby enabling the treatment of multiphase effects essential to liquid-fueled systems. Validation is performed against detailed computational fluid dynamics (CFD) simulations of a liquid-fueled can combustor, with parametric studies conducted across variations in inlet air temperature and fuel flow rate. Results show that the LFRN substantially reduces \ce{{NO\textsubscript{x}}} prediction errors relative to gaseous reactor networks while maintaining accurate outlet temperature predictions. A sensitivity analysis on the number of clusters demonstrates progressive convergence toward the CFD predictions with increasing network complexity. In terms of computational efficiency, the LFRN achieves runtimes of \textcolor{black}{$\mathcal{O}(10)$~s} on a single CPU core, representing speed-ups generally exceeding \textcolor{black}{1,000$\times$} compared to CFD. Overall, the findings demonstrate the potential of the LFRN as a computationally efficient reduced-order modeling tool that complements CFD to enable rapid emissions assessment and design-space exploration for liquid-fueled gas turbine combustors.
\end{abstract}


\section*{Novelty and Significance Statement}
\vspace{-5pt}
Current reactor network approaches are primarily suited to combustors in which the fuel is introduced in the gaseous phase. This work presents, to the authors’ knowledge, the first truly coupled reactor network framework for liquid-fueled combustors by introducing specialized reactors that capture spray evolution, including droplet breakup, heating, and evaporation, directly within the network. The significance of this study lies in enabling physics-based reduced-order modeling of multiphase liquid-fueled combustors, where conventional gaseous reactor networks are inadequate, while achieving predictions at orders of magnitude lower computational cost than CFD. This has important implications for efficient combustor design and optimization.

\section{Introduction}
\label{sec1}
\vspace{-7pt}
In recent decades, computational fluid dynamics (CFD), with varying levels of fidelity, has emerged as a powerful tool to aid in the design and optimization of turbulent combustors. At the high-fidelity end of the spectrum, direct numerical simulations (DNS) resolve all relevant temporal and spatial scales, thus offering detailed insights into combustor physics. DNS, however, involves extreme computational costs that render them infeasible for practical geometries and therefore, are not routinely used for design and optimization of practical combustors. When performing simulation-driven design and optimization, lower-fidelity models that offer better tractability, such as Reynolds-Averaged Navier Stokes (RANS), and more recently, Large Eddy Simulation (LES) are employed. RANS simulations with finite-rate chemistry provide detailed flow field information \cite{pope2000turbulent}, while LES approaches offer higher fidelity relative to RANS for turbulence-chemistry interactions \cite{pitsch2006large}. The Flamelet Generated Manifold (FGM) approach incorporates chemistry effects efficiently through precomputed lookup tables \cite{fiorina2010modelling}. These solvers can be coupled to an optimization algorithm (e.g., a genetic algorithm) and run over successive generations to virtually test a range of combustor configurations and ultimately discover promising designs over many design generations. However, in spite of their efficiency advantages over DNS, RANS and LES simulations still incur significant computational costs. A single RANS simulation of gas turbine combustors can take as much as $\mathcal{O} (\textrm{days})$ on $\mathcal{O} (10-100)$ processors to complete \cite{modest2013radiative}, depending on the geometry and simulation parameters. This computational burden prolongs the design cycle and thus motivates the development of even more efficient modeling tools that can be used for optimization and control.

One such framework is the chemical react{or} network (CRN) approach. In this framework, the three-dimensional combustor domain is divided into homogeneous zones, each representing distinct physical and chemical processes. Each zone is treated as an ideal reactor type (e.g., perfectly stirred reactor (PSR), plug flow reactor (PFR), or partially stirred reactor (PaSR)), depending on the dominant flow and combustion characteristics. The PSR assumes perfect mixing in highly turbulent regions, while PFR {is} well suited to flow zones with a strongly dominant flow direction \cite{turns2011introduction}. Previous studies have shown that CRNs can achieve computational speed-ups of several orders of magnitude while maintaining reasonable accuracy within restricted parameter ranges for gaseous combustion systems \cite{fichet2010reactor,kaluri2018realtime}.

A number of studies have explored the use of reactor networks to provide rapid estimates of various combustion quantities. 
Fichet et al. \cite{fichet2010reactor} developed CRN models for predicting {NO\textsubscript{x}} emissions in gas turbines, demonstrating good agreement with measured data in terms of quantitative levels and qualitative trends for gas turbine flame tubes operating at elevated pressures. 
The CRN approach by De Toni et al. \cite{DeToni2013} involved predicting {NO\textsubscript{x}} emissions in an industrial natural gas burner by extracting reactor configurations from experimental and CFD simulation results, with careful adjustment of reactor volumes and flow splits based on characteristic temperatures to reproduce experimental {NO\textsubscript{x}} emission data.
Perpignan et al. \cite{Perpignan2019} developed a joint CFD-Chemical Reactor Network (CRN) approach for modeling flameless combustion pollutant emissions, employing a three-step methodology that combines CFD simulations with simplified chemistry, automated clustering of computational cells into ideal reactors based on user-defined criteria, and a detailed chemical mechanism in the CRN.
Trespi et al. \cite{Trespi2021} presented the NetSMOKE solver within the OpenSMOKE++ suite for oxy-fuel combustion applications, enabling efficient chemical reactor network modeling with detailed kinetic mechanisms for novel combustor technologies, but found that computational cost increases significantly with the number of ideal reactors required. 

Savarese et al. \cite{Savarese2023} developed an automated methodology for generating CRNs from CFD simulations using machine learning clustering algorithms. The authors employed $k$-means clustering combined with graph scanning techniques to post-process CFD data of a MILD-capable semi-industrial furnace operating with \ce{CH4}-\ce{H2} fuel mixtures. Their approach successfully predicted {NO\textsubscript{x}} emissions across various fuel compositions and air injector configurations while achieving significant computational savings compared to full-order CFD simulations.
In another closely-related study, Savarese et al. \cite{Savarese2024} also developed a model-to-model Bayesian calibration framework for CRNs using data obtained from the CFD simulation of an ammonia-fueled multistage combustor.
Their approach successfully calibrated a compact 5-reactor CRN model against 15 RANS simulations, achieving mean {NO\textsubscript{x}} predictions with 4.2\% average relative error and enabling reliable optimization under uncertainty to identify optimal operating conditions that minimize {NO\textsubscript{x}} emissions.
Villette et al. \cite{Villette2024} developed a simplified three-element CRN for aeroengine combustion chambers, demonstrating accurate predictions of {NO\textsubscript{x}} emissions, CO concentrations, unburnt hydrocarbons, and combustion efficiency across various power settings.
Dübal et al. \cite{Dubal2024} extended the CRN framework to solid fuel combustion under oxy-fuel atmospheres, developing novel solid-gas plug flow reactors to predict CO formation, \ce{{NO\textsubscript{x}}}, \ce{{SO\textsubscript{x}}}, and aromatic pollutants. This study also highlighted the challenges many CRN approaches face with the coupling of solid and gas phase processes, which motivated the development of specialized reactor types to increase modeling fidelity. Du et al. \cite{du2014equivalent} applied a CFD-equivalent reactor network and used it to examine the effect of gasification temperature and equivalence ratio on a pilot-scale canonical spouted bed's performance.

The present study follows a similar paradigm, extending the CRN modeling framework to combustors where the fuel is introduced in the form of a liquid spray -- an approach we refer to as a liquid-fueled reactor network (LFRN). Applying CRNs to such systems presents unique challenges, as spray dynamics must be resolved. Traditional CRN methods assume homogeneous gas phase conditions, which are inadequate for capturing non-uniform fuel distribution and evaporation phenomena. The integration of spray dynamics into reactor network frameworks remains relatively underexplored and represents a significant gap in reduced-order modeling for liquid-fueled systems. The most related study is by Xu et al. \cite{xu2013procedure}, where a CRN approach was developed for capturing {NO\textsubscript{x}} emission in a lean premixed prevaporized combustor. However, their method did not explicitly incorporate spray dynamics within the reactor network. Instead, semi-analytical correlations and evaporation models were applied in a preprocessing step external to the CRN to compute the degree of evaporation, which was then used to define CRN boundary conditions and reactor partitioning. In contrast, the present work develops specialized reactors that directly capture spray-related physics (including breakup, droplet heating, and evaporation) and embeds these within the CRN framework. The modeling strategy also employs automated domain partitioning, mass balance enforcement, and optimization procedures for CRN calibration.

The remainder of this paper is organized as follows. Section \ref{sec:cfd_methodology} introduces the CFD setup and the development of the reactor network, including the mathematical models for the specialized reactors. Section \ref{sec:cfd_results} presents and analyzes the results, validating the LFRN approach through parametric studies benchmarked against full-order CFD simulations, including a discussion of computational time savings. Finally, Section \ref{sec:conclusions} provides concluding remarks.

\section{Methodology}
\label{sec:methodology}

\subsection{CFD Setup and Computational Domain}
\label{sec:cfd_methodology}
\vspace{-10pt}
\begin{figure}[!h]
    \centering
    \includegraphics[width=0.45\textwidth]{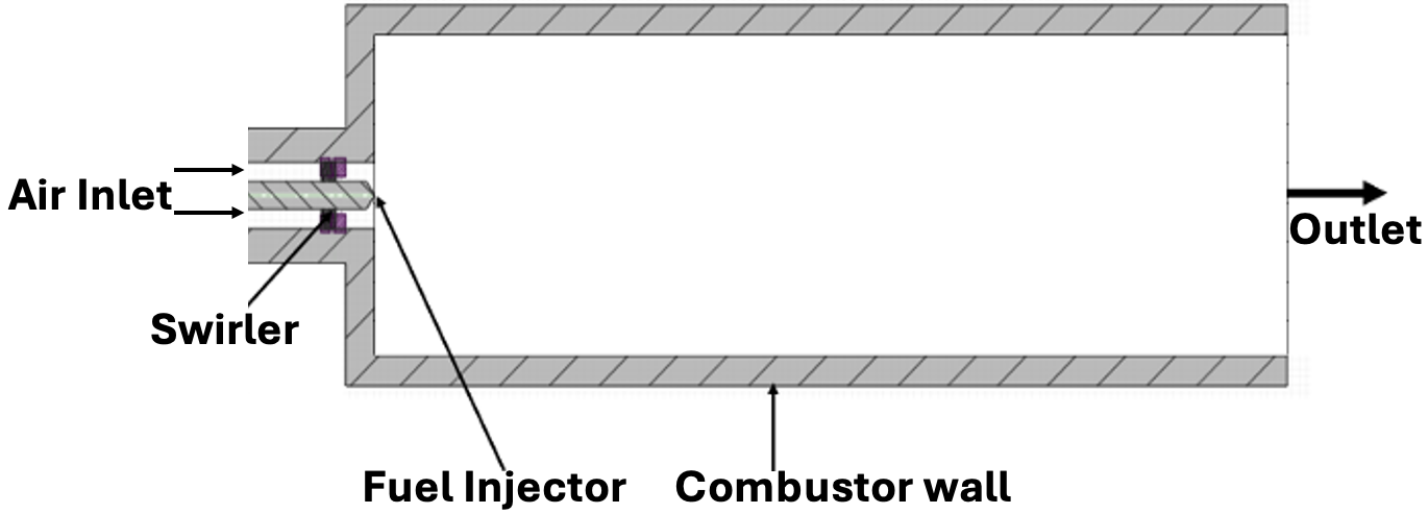}
    \caption{Schematic of the liquid-fueled swirl-stabilized combustor, illustrating the central fuel injector, surrounding swirled air inlet, and can-shaped combustion chamber.}
    \label{fig:geometry}
\end{figure}

The CFD simulations were performed using the CONVERGE CFD solver version 4.1 \cite{converge2025} with a Reynolds-Averaged Navier-Stokes (RANS) approach. The computational domain represents a swirl-stabilized combustor with geometric features shown in Fig.~\ref{fig:geometry}. The main combustion domain is can-shaped with a radius of 0.136 m and a height of 0.81 m. Air enters the combustor at 300 K through an annular inlet and passes through a swirler. The combustor geometry incorporates a central fuel injector with a diameter of 250 $\mu$m surrounded by the swirl-stabilized air inlet, designed to promote mixing and flame stabilization through the generation of a central recirculation zone. After spray injection into the combustion chamber, breakup, atomization, heating, and evaporation occur, followed by mixing with the surrounding gases and ignition. The domain extends from the fuel injection point through the combustion chamber to the exhaust plane, capturing the complete spray evolution and combustion processes. 

RANS turbulence closure was achieved using the standard $k$-$\varepsilon$ model \cite{launder1974application}, while the Flamelet Generated Manifold (FGM) approach \cite{van2000flamelet} was employed to circumvent the need to solve for complex chemistry within the CFD solver, thus promoting computational tractability. Specifically, we employ the FGM methodology to precompute detailed chemistry solutions for laminar flamelets under various conditions and store the results in a lookup table parameterized by progress variable $c$ and mixture fraction $Z$. During the table generation, the flamelet equations are solved in mixture fraction space \cite{peters2000turbulent}:
{
\setlength{\abovedisplayskip}{6pt}
\setlength{\belowdisplayskip}{6pt}
\begin{equation}
\rho\frac{\partial Y_i}{\partial t} = \frac{1}{2}\rho\chi\frac{\partial^2 Y_i}{\partial Z^2} + \dot{\omega}_i
\end{equation}
\begin{equation}
\rho c_p\frac{\partial T}{\partial t} = \frac{1}{2}\rho c_p \chi\frac{\partial^2 T}{\partial Z^2} - \sum_i h_i\dot{\omega}_i
\end{equation}
}
where the subscript $i$ denotes the $i^{\mathrm{th}}$ species, $Y_i$ is the species mass fraction, $T$ is the temperature, $Z$ is the mixture fraction, $\chi$ is the scalar dissipation rate, $\dot{\omega}_i$ is the chemical source term, $c_p$ is the specific heat at constant pressure, and $h_i$ is the specific enthalpy. 

The progress variable was defined as a linear combination of selected species as \textcolor{black}{$Y_c = Y_{CO_2} + Y_{CO}$ and the normalized progress variable as $c =  Y_c/Y_{eq}$} the {sub}script "eq" denotes equilibrium \textcolor{black}{values, computed at respective values of $Z$}. The transport equations for the progress variable and mixture fraction are solved in the CFD simulation using the following partial differential equations:
{
\setlength{\abovedisplayskip}{6pt}
\setlength{\belowdisplayskip}{6pt}
\begin{equation}
{\color{black}
\partial_t(\bar{\rho}\widetilde{c})
+
\nabla\cdot(\bar{\rho}\widetilde{\mathbf{u}}\widetilde{c})
=
\nabla\cdot(\bar{\rho}D_c\nabla\widetilde{c})
+
\bar{\rho}\widetilde{\dot{\omega}}_{c}
}
\end{equation}
\begin{equation}
{\color{black}
    \partial_t(\bar{\rho}\widetilde{Z})
+
\nabla\cdot(\bar{\rho}\widetilde{\mathbf{u}}\widetilde{Z})
=
\nabla\cdot(\bar{\rho}D_Z\nabla\widetilde{Z})
+
\bar{S}_{\mathrm{spray}}}
\end{equation}
\begin{equation}
    \begin{aligned}
    {\color{black}
\partial_t(\bar{\rho}\widetilde{Z''^2})
+
\nabla\cdot(\bar{\rho}\widetilde{\mathbf{u}}\widetilde{Z''^2})}
&=
{\color{black}
\nabla\cdot(\bar{\rho}D_t\nabla\widetilde{Z''^2})} \\
&{\color{black}\quad
+ 2\bar{\rho}D_t|\nabla\widetilde{Z}|^2
-
\bar{\rho}\widetilde{\chi}.}
\end{aligned}
\end{equation}
}
\textcolor{black}{where $\partial_t(\cdot)$ denotes $\partial(\cdot)/\partial t$.} \textcolor{black}{$D_c = D_Z = D+D_t$, with $D$ and $D_t = {\nu}_t/Sc_t$ representing the molecular and turbulent diffusion coefficients, respectively, where $Sc_t$ is the turbulent Schmidt number} \textcolor{black}{taken as $Sc_t = 0.78$}. \textcolor{black}{The scalar dissipation rate is modeled as $\widetilde{\chi} = 2.0 (\tilde{\epsilon}/\widetilde{k})\tilde{Z''^2}$, where $\widetilde{\epsilon}$ is the turbulent dissipation rate and $\widetilde{k}$ is the turbulent kinetic energy}. $\widetilde{\dot{\omega}}_c$ is the progress variable source term obtained from the flamelet library \cite{fiorina2005modelling}. \textcolor{black}{During table construction, turbulence-chemistry interaction is captured using a presumed $\beta$-PDF for $Z$, parameterized by $\widetilde{Z}$ and $\widetilde{Z''^{2}}$, so that the resulting state variables depend on $\widetilde{Z}$, $\widetilde{Z''^2}$, and $\widetilde{c}$. Because NO\textsubscript{x} chemistry evolves on timescales considerably longer than those of the main flame chemistry, a separate progress variable, $c_{\ce{NO}} = Y_{\ce{NO}}$, was transported for the lookup of $Y_{\ce{NO}}$ and $Y_{\ce{NO2}}$.}

The liquid fuel spray was modeled using Lagrangian tracking with the discrete droplet method. The Taylor Analogy Breakup (TAB)~\cite{orourke1987} model was employed to simulate both primary and secondary breakup processes. The Frossling correlation ~\cite{amsden1989} was applied to describe the droplet evaporation process with the droplet size distribution characterized by a Rosin-Rammler distribution \cite{rosinrammler1933}.

\subsection{Reactor Network Development}
\label{sec:dev_rn}
\begin{figure*}[htbp]
    \centering
    \includegraphics[width=0.62\textwidth]{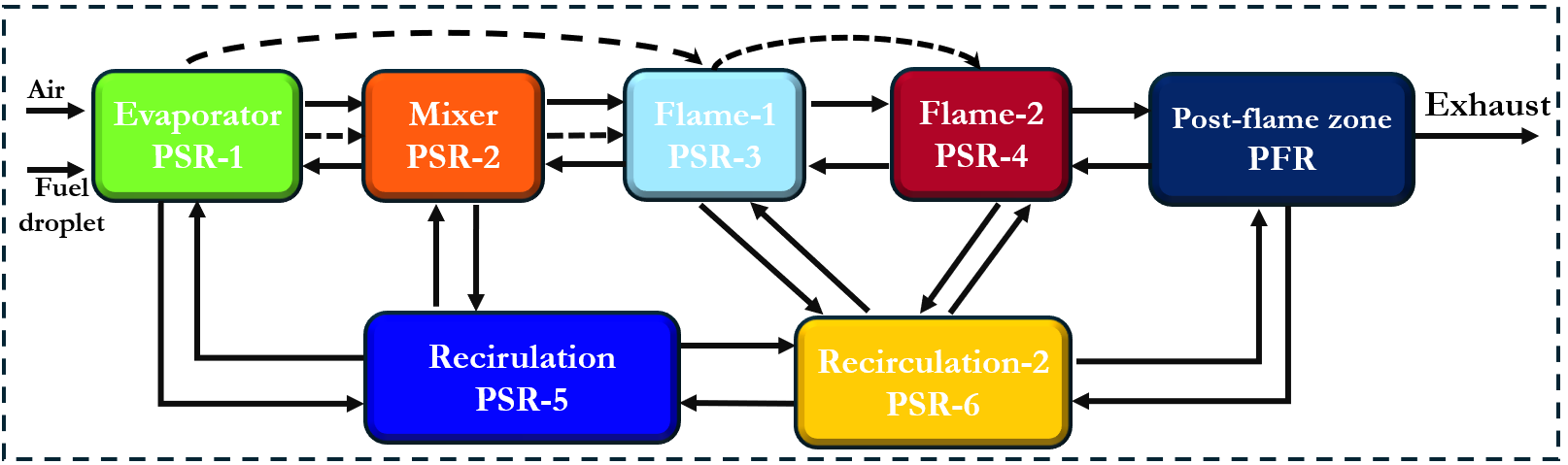}
    \caption{Seven-cluster reactor network configuration. \textcolor{black}{Solid lines represent gas-phase mass transfer, while the dashed lines denote the transport of droplets.}}
    \label{fig:network}
\end{figure*}

\vspace{-4pt}
In this section, we introduce the modeling framework for the reactor network. In contrast to the CFD modeling that involves the use of millions of computational cells, in the reactor network approach, the physical space within the combustor is divided into zones, with the evolution of the thermochemical state in each reactor governed by a simplified transport formulation. Conventional reactor types are employed for some zones, including perfectly stirred reactors (PSRs), where the fluid is assumed to be perfectly mixed with no spatial gradients within the reactor volume, and plug flow reactors (PFRs), where all gradients are confined to the axial (flow) direction and absent in the transverse directions. To capture the distinctive effects of liquid fuel, two additional specialized reactor types are introduced: (1) an evaporator/breakup zone to represent droplet breakup with partial evaporation, and (2) a mixer to account for continued droplet heating, evaporation, and subsequent gas phase fuel–air mixing prior to combustion. An illustration of the reactor network configuration and the coupling between reactors is shown in Fig.~\ref{fig:network}. 



The air and liquid fuel mixture is introduced into the reactor network via the evaporator. The role of the evaporator is to capture the initial spray breakup and evaporation from the liquid to the gas phase. The liquid fuel treatment incorporates simplified spray models that capture essential physics without full Lagrangian complexity. Consistent with the assumptions of spatially uniform reactors such as PSRs, we \textcolor{black}{begin by representing} the spray with a single representative droplet \textcolor{black}{value} rather than tracking a full droplet size distribution. \textcolor{black}{As droplets are transported through the reactor network, however, a downstream reactor may receive droplets from multiple upstream reactors. In such cases, the downstream reactor retains separate droplet classes, each associated with its own upstream history and current size, temperature, and evaporation state. These classes are evolved independently using the droplet heating and evaporation equations, and the total evaporation source supplied to the reactor is obtained by summing over all resident droplet classes.}
The initial number of droplets introduced into the evaporator was calculated based on the fuel flow rate and the radius of the nozzle.
The initial droplet radius is assumed to equal the nozzle radius, so that $r_o = r_{nozzle}$.
The number of droplets introduced per unit time, $\dot{N}_d$, is determined from the mass flow rate as

\begin{equation}
\dot{N}_d = \frac{\dot{m}_{fuel}}{m_d} = \frac{\dot{m}_{fuel}}{\rho_l \frac{4\pi}{3} r_o^3},
\end{equation}

where $r_o$ is the initial droplet radius, $m_d$ is the mass of a single droplet, $\rho_l$ is the liquid fuel density, $\dot{m}_{fuel}$ is the mass flow rate of fuel.

The fuel is transported from the evaporator to the mixer, where the representative droplet undergoes further heating and evaporation. The implicit assumption here is that the breakup happens on a much faster timescale compared to evaporation, so that both processes can be decoupled. We provide a justification for this assumption later in Section~\ref{sec:timescale_analysis}. In general, the breakup process modifies the droplet diameter to produce a larger number of smaller-sized droplets, while the evaporation process captures the mass transfer from the liquid to the gas phase. The evaporated mass is introduced as a source term for the fuel species in the corresponding reactors where the evaporating droplet is present. Droplets that fail to fully evaporate at the prescribed residence time for a given reactor are convected to downstream reactors, where they continue to evaporate, analogous to the 3-dimensional CFD formulation. Consistent with the homogeneous reactor network assumption, perfectly stirred reactors are used, such that complete mixing is assumed once evaporation occurs.
The mathematical models used for droplet breakup, evaporation, and heating are described in the following subsections.

\subsubsection{Droplet breakup}
\vspace{-3pt}
Similar to the CFD modeling, the breakup of liquid droplets is described using the TAB approach developed by O'Rourke and Amsden~\cite{orourke1987}. Only a summary is provided here. More details can be found in the original paper. The TAB approach models droplet deformation as a forced, damped harmonic oscillator with spring, damping, and forcing terms:
{
\setlength{\abovedisplayskip}{6pt}
\setlength{\belowdisplayskip}{6pt}
\begin{equation}
\ddot{x} + \frac{c}{m}\dot{x} + \frac{k}{m}x = \frac{F}{m},
\label{eq:tab_basic}
\end{equation}
}
where $x$ represents the deviation of the droplet’s equatorial position from its undisturbed spherical state. The model coefficients, derived from Taylor’s analogy, are expressed as
{
\setlength{\abovedisplayskip}{6pt}
\setlength{\belowdisplayskip}{6pt}
\begin{equation}
\frac{F}{m}=C_f\frac{\rho_g u_d^2}{\rho_l r_0},
\quad
\frac{k}{m}=C_k\frac{\sigma}{\rho_l r_0^3},
\quad
\frac{c}{m}=C_d\frac{\mu_l}{\rho_l r_0^2},
\label{eq:tab_coeffs}
\end{equation}
}
where $\rho_l$ and $\rho_g$ are the liquid and gas densities, $u_d$ is the relative droplet–gas velocity, $r_0$ is the initial droplet radius, $\sigma$ is the surface tension, and $\mu_l$ is the liquid viscosity. The empirically determined dimensionless parameters are $C_k = 8$, $C_f = 1/3$, {$C_d = 5$}, and $C_b = 1/2$. Introducing the nondimensional displacement $y = x / (C_b r_0)$ and substituting the coefficient relationships from Eq.~\ref{eq:tab_coeffs}, the governing equation becomes
{
\setlength{\abovedisplayskip}{6pt}
\setlength{\belowdisplayskip}{6pt}
\begin{equation}
\frac{d^2y}{dt^2} = \frac{C_f \rho_g u_d^2}{C_b \rho_l r_0^2} - \frac{C_k \sigma}{\rho_l r_0^3} y - \frac{C_d \mu_l}{\rho_l r_0^2} \frac{dy}{dt}.
\label{eq:tab_nondimensional}
\end{equation}
}


Breakup is determined by the oscillation amplitude, which depends on the droplet oscillation frequency $\omega$ and the Weber number $We_g$, defined as:
{
\setlength{\abovedisplayskip}{6pt}
\setlength{\belowdisplayskip}{6pt}
\begin{align}
\omega^2 &= C_k \frac{\sigma}{\rho_l r_0^3} - \left(\frac{C_d \mu_l}{2 \rho_l r_0^2}\right)^2, \label{eq:omega} \\
\text{We}_g &= \frac{\rho_g u_d^2 r_0}{\sigma}. \label{eq:weber}
\end{align}
From these, we compute the oscillation amplitude as:
\begin{equation}
A = \sqrt{\left(y - \frac{C_f}{C_k C_b} \text{We}_g \right)^2 + \left(\frac{\dot{y}}{\omega}\right)^2}
\label{eq:amplitude}
\end{equation}
}

Breakup is inhibited when $A + \text{We}_g \leq 1.0$, while breakup becomes possible when $A + \text{We}_g > 1.0$. Upon breakup, the resulting droplet size is determined by the O'Rourke and Amsden expression:
{
\setlength{\abovedisplayskip}{6pt}
\setlength{\belowdisplayskip}{6pt}
\begin{equation}
r = \frac{r_0}{1 + \frac{8K}{20}y^2 + \frac{\rho_l r_0^3}{\sigma}\dot{y}^2\left(\frac{6K-5}{120}\right)},
\label{eq:breakup_radius}
\end{equation}
}
where $K = 10/3$ and $r_0$ represents the pre-fragmentation droplet radius.

\subsubsection{Droplet evaporation}
In addition to liquid fuel breakup, the LFRN also captures the liquid-to-vapor conversion process, based on the Frossling correlation~\cite{amsden1989}. The temporal evolution of the droplets is described by the following ordinary differential equation:
{
\setlength{\abovedisplayskip}{6pt}
\setlength{\belowdisplayskip}{6pt}
\begin{equation}
\frac{dr}{dt} = -\frac{\rho_g D}{2 \rho_l r} B_d \text{Sh}_d.
\label{eq:frossling}
\end{equation}
}
In this expression, $r$ is the post-breakup droplet radius from Eq.~\ref{eq:breakup_radius},  $\rho_g$ is the gas density, $D$ represents the vapor-air mass diffusivity, and $\text{Sh}_d$ is the Sherwood number. The driving force parameter $B_d$ is defined as:
{
\setlength{\abovedisplayskip}{6pt}
\setlength{\belowdisplayskip}{6pt}
\begin{equation}
B_d = \frac{Y_s - Y_{\infty}}{1 - Y_s},
\label{eq:bd_parameter}
\end{equation}
}
where $Y_s$ represents the mass fraction of the fuel vapor at the droplet surface (i.e., $r = r_0$), while $Y_{\infty}$ represents the fuel vapor mass fraction in a bulk of the gas phase, far away from the droplet (i.e., $r = \infty$). This driving force parameter is proportional to the difference in fuel vapor mass fraction between the droplet surface and the far-field gas. When $Y_s = Y_{\infty}$, the equilibrium condition is reached and no net mass transfer occurs. The Sherwood number incorporates both molecular and convective transport effects:
{
\setlength{\abovedisplayskip}{6pt}
\setlength{\belowdisplayskip}{6pt}
\begin{equation}
\text{Sh}_d = \left(2.0 + 0.6\text{Re}_d^{1/2}\text{Sc}^{1/3}\right) \frac{\ln(1 + B_d)}{B_d},
\label{eq:sherwood}
\end{equation}
}
where $Re_d$ is the Reynolds number characterizing droplet motion and is described as $\text{Re}_d = \rho_g u_d d/\mu_{air}$, while the Schmidt number describing momentum-to-mass transport analogy is $\text{Sc} = \mu_{air}/\rho_g D.$ The air viscosity $\mu_{air}$ is evaluated at the reference temperature:
{
\setlength{\abovedisplayskip}{6pt}
\setlength{\belowdisplayskip}{6pt}
\begin{equation}
\bar{T} = \frac{T_{gas} + 2T_d}{3},
\label{eq:temperature_avg}
\end{equation}
}
where $T_{gas}$ and $T_d$ are the gas and droplet temperatures, respectively. Finally, the mass diffusivity is modeled using the temperature-dependent correlation:
{
\setlength{\abovedisplayskip}{6pt}
\setlength{\belowdisplayskip}{6pt}
\begin{equation}
\rho_g D = 1.293 D_0 \left(\frac{\bar{T}}{273}\right)^{n_D-1}, 
\label{eq:diffusivity}
\end{equation}
}
with $D_0$ and $n_D$ as experimentally determined constants. For Jet-A fuel, $D_0 = 4.16 \times 10^{-6}$ and $n_D = 1.6$.

\subsubsection{Thermal Evolution: Uniform Temperature Approach}
For the reactor network model, we assume that the droplets have a uniform internal temperature distribution \textcolor{black}{following the infinite liquid-conductivity model, in which the droplet temperature varies in time but remains spatially uniform \cite{sirignano2010fluid, sazhin2014droplets}. This is the same droplet-heating approximation used in the CFD spray calculation. Future work will explore more detailed models, such as spherically symmetric transient droplet-heating formulations or finite-conductivity models}. Based on energy conservation considerations, the heat gained by the droplet is balanced with the increase in sensible enthalpy and loss of enthalpy due to vaporization, yielding ~\cite{amsden1989}:
{
\setlength{\abovedisplayskip}{6pt}
\setlength{\belowdisplayskip}{6pt}
\begin{equation}
c_l m_d \frac{dT_d}{dt} =  \frac{A_d\beta_{spray} \text{Nu}_d k_\textrm{air} (T_{gas} - T_d)}{2r} + \frac{dm_d}{dt} l_{vap},
\label{eq:energy_balance}
\end{equation}
}
where $c_l$ is the liquid specific heat capacity, $A_d$ is the surface area of the droplet, $T_d$ is the droplet temperature, $m_d$ is the droplet mass, $r$ is the droplet radius, and $l_{vap}$ is the temperature-dependent latent heat of vaporization. The heat transfer coefficient is scaled by the user-defined parameter $\beta_{spray}$, while $k_{air}$ represents the thermal conductivity evaluated at $\bar{T}$. The Nusselt number, $\text{Nu}_d$, is defined as
{
\setlength{\abovedisplayskip}{6pt}
\setlength{\belowdisplayskip}{6pt}
\begin{equation}
\text{Nu}_d = \left(2.0 + 0.6 \text{Re}_d^{1/2} \text{Pr}_d^{1/3}\right) \frac{\ln(1 + B_d)}{B_d},
\label{eq:nusselt}
\end{equation}
}
where $\text{Re}_d$ is the droplet Reynolds number, $\text{Pr}_d$ is the Prandtl number, and $B_d$ is the driving-force parameter defined in Eq.~\ref{eq:bd_parameter}. The quantities $A_d$, $r$, and $m_d$ are taken from the previous time step during the numerical integration.

\subsubsection{Integrated Process Coupling}

These three phenomena, namely breakup, heating and evaporation, are allowed to modify the liquid droplet. In the evaporator, using the TAB methodology described, we first assess breakup potential using the amplitude criterion (Eq.~\ref{eq:amplitude}). When $A + \text{We}_g > 1.0$, breakup occurs and produces a new characteristic droplet radius $r$ via Eq.~\ref{eq:breakup_radius}. The droplet radius obtained after breakup serves as the initial condition for subsequent mass transfer calculations that include evaporation and heating. Unlike breakup and evaporation, which are decoupled, the evaporation and heating processes are coupled. The uniform temperature model (Eq.~\ref{eq:energy_balance}) evolves the droplet temperature, which influences the vapor pressure, and therefore, impacts the fuel vapor mass fraction at the surface via the following relationship ~\cite{amsden1989}:
{
\setlength{\abovedisplayskip}{6pt}
\setlength{\belowdisplayskip}{6pt}
\begin{equation}
Y_F = \frac{MW_F}{MW_F + MW_{mix} \left(\frac{p_{g}}{p_v} - 1\right)},
\label{eq:vapor_mass_fraction}
\end{equation}
}
where $MW_{mix}$ is the vapor-free mixture molecular weight, $p_{g}$ is the gas pressure, and $p_v = p_v(T)$ is the temperature-dependent vapor pressure. Therefore, according to Eq.~\ref{eq:vapor_mass_fraction}, since the vapor pressure increases with temperature, higher droplet temperatures will result in a higher fuel mass fraction at the droplet surface, $Y_F$, which in turn drives an accelerated rate of evaporation via the driving force parameter, $B_d$ (Eqs. \ref{eq:bd_parameter} and \ref{eq:frossling}). \textcolor{black}{Vapor pressure values used in this study were tabulated from experimental data \cite{edwards2017reference}.}

\subsubsection{$k$-means Clustering and Optimization} \label{sec:clustering_opt}
\noindent\textit{$k$-means Clustering}  

\noindent The development of the reactor network begins with automated domain partitioning using $k$-means clustering. Thermochemical and flow-field data are extracted from three-dimensional CFD simulations under representative operating conditions. For each cell, we form a feature vector consisting of temperature and equivalence ratio. For the $i^{\mathrm{th}}$ cell, this vector is defined as; $\bm{\theta}_i = [T_i, \, \phi_i]$. \textcolor{black}{$\phi$ was chosen to capture key regions of the combustor with distinct mixture states, such as regions near the nozzle where fuel evaporation can produce locally rich mixtures, while temperature was selected as a marker of reactivity}. The complete dataset is assembled into a matrix, $\bm{\Theta} = [\bm{\theta}_1^T, \bm{\theta}_2^T, \ldots, \bm{\theta}_{n_o}^T]^T$, where $n_o$ is the number of observations (i.e., CFD cells). Using this dataset, the domain is partitioned into $n_r$ clusters, each corresponding to a homogeneous reactor zone. The $k$-means algorithm achieves this by minimizing the within-cluster sum of squares: 
{
\setlength{\abovedisplayskip}{5pt}
\setlength{\belowdisplayskip}{5pt}
\begin{equation}\label{eq:kmeans}
\mathcal{J}^{k\text{-means}} = \sum_{j=1}^{n_r} \sum_{\bm{\theta}_i \in C_j} \|\bm{\theta}_i - \mu_j\|^2,
\end{equation}
}
where $C_j$ is the set of points assigned to the $j$th cluster and $\mu_j$ is its centroid. Equation~\ref{eq:kmeans} implies that the centroids are selected to minimize the sum of Euclidean distances from all the points to their assigned centroids. The algorithm proceeds iteratively by alternately updating cluster centroids based on the current point assignments and then reassigning points to the nearest updated centroid until convergence.  

\vspace{10pt}
\noindent\textit{Mass Correction}  

After the CFD solution is obtained, the thermochemical and flow-field data are extracted from the Adaptive Mesh Refinement (AMR) grid. Because the AMR grid is unstructured, the extracted fields are first interpolated onto a structured auxiliary mesh. The mass flux through each face is then computed as $\dot{m}_f = \rho_f \left(\mathbf{u}_f \cdot \mathbf{n}_f\right) A_f, $ where $\rho_f$ and $\mathbf{u}_f$ are the face-interpolated density and velocity, $\mathbf{n}_f$ is the outward unit normal vector, and $A_f$ is the face area. These interpolation and post-processing operations can introduce small mass-conservation errors. 
{Table~\ref{tab:reactor_imbalance} reports the raw imbalance, $\varepsilon =\dot{m}_{\mathrm{in}}-\dot{m}_{\mathrm{out}}$, and the relative imbalance, $\varepsilon/\dot{m}_{\mathrm{in}}\times100\%$ before correction.}


\begin{table}[htbp]
\centering
\footnotesize
\color{black}
\setlength{\tabcolsep}{3pt}
\caption{{Initial reactor mass flow rates before correction.}}
\label{tab:reactor_imbalance}
\begin{tabular}{@{}lrrrc@{}}
\toprule
Reactor &
$\dot{m}_{\mathrm{in}}$ (g/s) &
$\dot{m}_{\mathrm{out}}$ (g/s) &
$\varepsilon$ (g/s) &
$\varepsilon/\dot{m}_{\mathrm{in}}$ (\%) \\
\midrule
PFR   & 123.67 & 123.44 & $+0.23$ & $+0.19$ \\
PSR-4 & 202.21 & 203.51 & $-1.30$ & $-0.64$ \\
PSR-2 &  40.96 &  39.95 & $+1.01$ & $+2.46$ \\
PSR-1 & 127.54 & 130.70 & $-3.15$ & $-2.47$ \\
PSR-3 & 129.07 & 129.90 & $-0.82$ & $-0.64$ \\
PSR-6 &   5.80 &   6.34 & $-0.54$ & $-9.31$ \\
PSR-5 & 155.76 & 151.18 & $+4.58$ & $+2.94$ \\
\bottomrule
\end{tabular}
\end{table}

{To balance the mass, the reactor-to-reactor mass-flow matrix is assembled by identifying all faces shared by pairs of clustered regions. For a reactor network with $n_r$ reactors, we define $\mathbf{M} \in \mathbb{R}^{n_r \times n_r},$
where $m_{ij}$ denotes the total mass flow rate from reactor $i$ to reactor $j$. This quantity is obtained by summing the outward mass fluxes over all faces separating reactors $i$ and $j$ with the face normal taken outward from reactor $i$, {while ignoring regions of the boundary with flow from $j$ to $i$. Elements of $\mathbf{M}$ involving unconnected reactors are set to $m_{ij} = 0$}. For each reactor $i$, mass conservation requires the total mass entering the reactor to equal the total mass leaving it. This condition is written as}

{
\setlength{\abovedisplayskip}{6pt}
\setlength{\belowdisplayskip}{6pt}
{
\begin{equation}
{\color{black}
\dot{m}_{i,\mathrm{in}}+\sum_{j=1}^{n_r}\dot{m}_{ji}=\dot{m}_{i,\mathrm{out}}+\sum_{j=1}^{n_r}\dot{m}_{ij}, 
}
\label{eq:mass_balance}
\end{equation}
}
}
{
where $\dot{m}_{i,\mathrm{in}}$ and $\dot{m}_{i,\mathrm{out}}$ denote prescribed external inlet and outlet contributions, respectively, for reactors in contact with the domain boundaries.}
{
{To enforce this, positive mass-flow rates, as obtained from the CFD data, are selected and flattened} into a vector $\mathbf{z} = \left[m_{ij} \;:\; m_{ij} > 0 \right]^T$. We can correct this vector by adding a small adjustment, such that $\mathbf{z}^{*}=\mathbf{z}+\Delta\mathbf{z}$. The correction $\Delta\mathbf{z}$ is chosen to be the smallest change that restores mass conservation in every reactor. This is implemented as a minimum-change least-squares correction, $\Delta\mathbf{z}^{*}=\arg\min_{\Delta\mathbf{z}}\|\Delta\mathbf{z}\|_2^2$, subject to the mass-balance in Eq. \ref{eq:mass_balance} being satisfied for every reactor. In practice, this step enforces mass conservation while minimally perturbing the optimizer-proposed mass-flow rates.}


\noindent\textit{Optimization Procedure}  

{Furthermore, because the reactor network represents a highly reduced approximation of the original CFD solution, additional calibration is generally required to reproduce key quantities of interest. This adjustment modifies the bulk reactor residence time through $\tau_i=\bar{\rho}_i V_i/\dot{m}_i$. However, since each reactor represents a coarse-grained CFD region with spatially varying density, velocity, temperature, and composition, $\tau_i$ should be interpreted as an effective {reactor-scale residence time rather than the residence time associated with any individual CFD cell or fluid parcel.}}
The reactor network is calibrated by minimizing a weighted sum of relative errors in selected quantities of interest. The objective function is
{
\setlength{\abovedisplayskip}{6pt}
\setlength{\belowdisplayskip}{6pt}
\begin{equation}
    {\color{black}
    \mathcal{J}^{\mathrm{RN}}
    =
    \sum_{k=1}^{N_q}
    w_k
    \frac{
    \left|
    p_k^{\mathrm{RN}}(\boldsymbol{\theta})
    -
    p_k^{\mathrm{CFD}}
    \right|
    }{
    \left|
    p_k^{\mathrm{CFD}}
    \right|
    },}
\end{equation}
}
\textcolor{black}{where $\boldsymbol{\theta}$ denotes the complete set of optimization variables. The target vector is $\mathbf{p}=\left[ T_{\mathrm{out}}, Y_{\mathrm{NO,out}}, Y_{\mathrm{NO_2,out}}, \phi_m \right]^T ,$
where the subscript ``out'' denotes outlet quantities and $\phi_m$ denotes the equivalence ratio in the mixer reactor. The weights $w_k$ allow the calibration to emphasize selected targets. The weights are a user-choice which are set to $\mathbf{w} = \left[5,\;4,\;2,\;2 \right]^T$ for $T_{\mathrm{out}}$, $Y_{\mathrm{NO,out}}$, $Y_{\mathrm{NO_2,out}}$, and $\phi_m$.}

\textcolor{black}{The volume and spray-related variables are optimized only for reactors located along the spray path. We define the spray-path set as $\mathcal{S}$, which contains the sequence of reactors connecting the fuel nozzle to the reactor with the maximum temperature. For these reactors, the optimized volume vector is $\mathbf{v}_{\mathrm{opt}} = [V_i : i \in \mathcal{S}]^T$, where $V_i$ is the volume of reactor $i$. The spray path length $s_i$ represents the average distance traveled by the representative droplet within reactor $i$, and is optimized only for $i \in \mathcal{S}$. Thus, the spray-length vector is $\mathbf{s}_{\mathrm{opt}} = [s_i : i \in \mathcal{S}]^T$. The spray propagation fractions describe how liquid fuel leaving a reactor is distributed among downstream candidate reactors. Specifically, if spray leaving reactor $i \in \mathcal{S}$ can enter two candidate reactors $j$ and $k$, where both candidates share a mass-flow connection with reactor $i$ and also lie along the spray path, we optimize only one independent fraction, denoted $\eta_{ij}$. The fraction of spray sent from reactor $i$ to reactor $j$ is then $\eta_{ij}$, while the fraction sent to reactor $k$ is $1-\eta_{ij}$, thereby enforcing conservation of the outgoing spray fraction. The corresponding spray-fraction vector is written as $\boldsymbol{\eta}_{\mathrm{opt}} = [\eta_{ij} : i,j \in \mathcal{S}]^T$, where only independent fractions are included. The complete spray-variable vector is therefore $\mathbf{y}_{\mathrm{opt}} = [\mathbf{s}_{\mathrm{opt}}^T,\boldsymbol{\eta}_{\mathrm{opt}}^T]^T$.}

The mass-flow variables are selected separately through a sensitivity-based reduction procedure. A finite-difference perturbation is applied to each candidate mass-flow variable, and the resulting sensitivity of the reactor-network objective is denoted by $S_\ell$ for the $\ell$th candidate mass-flow rate. The mass-flow variables are then sorted in descending order of sensitivity, such that $S_{\hat{m}_1} \geq S_{\hat{m}_2} \geq \cdots \geq S_{\hat{m}_{n_z}}$, and the $n_r$ most sensitive variables are retained for optimization. The optimized mass-flow vector is therefore $\mathbf{z}_{\mathrm{opt}} = [\hat{m}_1,\hat{m}_2,\ldots,\hat{m}_{n_r}]^T$.

\begin{table}[H]
\centering
\footnotesize
\color{black}
\setlength{\tabcolsep}{4pt}
\caption{{Mass-balanced reactor inflows before optimization  ($\dot{m}_{\textrm{in}}$) and after optimization ($\dot{m}^{*}_{\textrm{in}}$). The relative change is computed as $(\dot{m}^{*}_{\textrm{in}}-\dot{m}_{\textrm{in}})/\dot{m}_{\textrm{in}}\times100\%$.}}
\label{tab:reactor_reconciled}
\begin{tabular}{@{}lrrc@{}}
\toprule
Reactor & $\dot{m}_{\textrm{in}}$ (g/s) & $\dot{m}^{*}_{\textrm{in}}$ (g/s) & Change (\%) \\
\midrule
PFR   & 123.44 &  93.99 & $-23.9$ \\
PSR-4 & 201.07 & 218.39 & $+8.6$  \\
PSR-2 &  38.73 &  33.59 & $-13.3$ \\
PSR-1 & 127.73 & 142.84 & $+11.8$ \\
PSR-3 & 129.3 & 162.63 & $+25.8$ \\
PSR-6 &   5.72 &   4.52 & $-21.0$ \\
PSR-5 & 152.51 & 112.15 & $-26.4$ \\
\bottomrule
\end{tabular}
\end{table}

Combining the selected mass-flow variables, reactor volumes, and spray variables gives the complete optimization vector $\boldsymbol{\theta} = [\mathbf{z}_{\mathrm{opt}}^T,\mathbf{v}_{\mathrm{opt}}^T,\mathbf{y}_{\mathrm{opt}}^T]^T$. The mass-flow variables and reactor volumes are allowed to vary by $\pm 30\%$ from their baseline values, while the spray path lengths are allowed to vary by $\pm 50\%$.
After each optimization update, the mass-flow variables proposed by the optimizer are corrected {to enforce mass conservation} before the reactor network is evaluated. This correction is necessary because independent changes to selected mass-flow rates can violate the mass balance of individual reactors. 

The final calibration problem is therefore to minimize $\mathcal{J}^{\mathrm{RN}}(\mathbf{z}^{*},\mathbf{v}_{\mathrm{opt}},\mathbf{y}_{\mathrm{opt}})$ subject to the prescribed bounds on the mass-flow rates, reactor volumes, spray path lengths, and spray propagation fractions, with $\mathbf{z}^{*}$ always satisfying the reactor mass balances. Because the objective function is nonlinear and may contain local minima, each calibration case is solved using both a genetic algorithm and the covariance matrix adaptation evolution strategy (CMA-ES)\cite{hansen2001cmaes}. The solution with the lower value of $\mathcal{J}^{\mathrm{RN}}$ is retained. If the reactor-network solver fails for a proposed parameter set, the candidate is assigned a penalty value of $\mathcal{J}^{\mathrm{RN}}=10^6$, which discourages further exploration of infeasible regions of the parameter space.

{Table~\ref{tab:reactor_reconciled} reports the change between the initial mass-balanced inflows and the final calibrated inflows returned by the optimizer.
Because $\tau_i = \bar{\rho}_i V_i / \dot{m}_i$, a fractional change $\Delta \dot{m}_i / \dot{m}_i$ produces, at fixed volume and mean density, a comparable inverse change in effective residence time. For example, the $-26.4\%$ change in the inlet mass flow for PSR-5 corresponds to an approximately $+35.9\%$ increase in $\tau_{\mathrm{PSR\text{-}5}}$, while a +25.8\% change in inlet mass flow for PSR-3 results in a -20.5\% change in $\tau_{\mathrm{PSR\text{-}3}}$. These residence-time adjustments are the physical mechanism through which the calibration influences the NO and NO$_2$ formation pathways predicted by the reactor network.}


\subsection{Computational Implementation}  
The computational mesh employed a uniform base grid size of \textcolor{black}{\SI{3.5}{\milli\meter}} with (AMR) down to \textcolor{black}{\SI{0.4375}{\milli\meter}} in regions with higher gradients, especially those close to the fuel injector, flame zones, and recirculation zones.  Using AMR, the mesh is dynamically refined or coarsened based on temperature gradients, velocity magnitude, and species concentration gradients to maintain adequate resolution of important spatial features with enhanced computational efficiency. The final mesh contained approximately 3.2 million cells after refinement, with enhanced resolution in regions of high scalar dissipation and turbulent mixing.

The reactor network equations were solved using the Cantera chemical kinetics library (v2.6.0) \cite{cantera2022}, coupled with a HyChem mechanism for Jet-A fuel combustion \cite{wang2018hychem1,xu2018hychem2}. The HyChem mechanism consists of 119 species and 841 reactions, with a seven-reaction fuel pyrolysis submodel combined with the USC Mech II mechanism (consisting of 111 species and 784 reactions) \cite{wang2007uscmech}. This physics-based approach decouples fuel pyrolysis from the oxidation of pyrolysis products: fuel decomposition is represented by experimentally constrained lumped reactions that yield key intermediates (primarily ethylene, methane, propene, butenes, benzene, and toluene), which are subsequently oxidized using detailed USC Mech II kinetics. This mechanism was selected because it offers an effective balance between chemical fidelity and computational tractability for the present reactor-network application. In particular, accurate \ce{{NO\textsubscript{x}}} prediction requires resolving thermal, prompt, and \ce{N2O}-intermediate pathways, all of which depend on detailed oxidation chemistry that the HyChem framework captures reliably.

\section{Results and Discussion}
\label{sec:results}

\subsection{Comparing Droplet Breakup and Evaporation Timescales} \label{sec:timescale_analysis}

\vspace{-8pt}
\begin{figure}[H]
    \centering
    \includegraphics[width=0.3\textwidth]{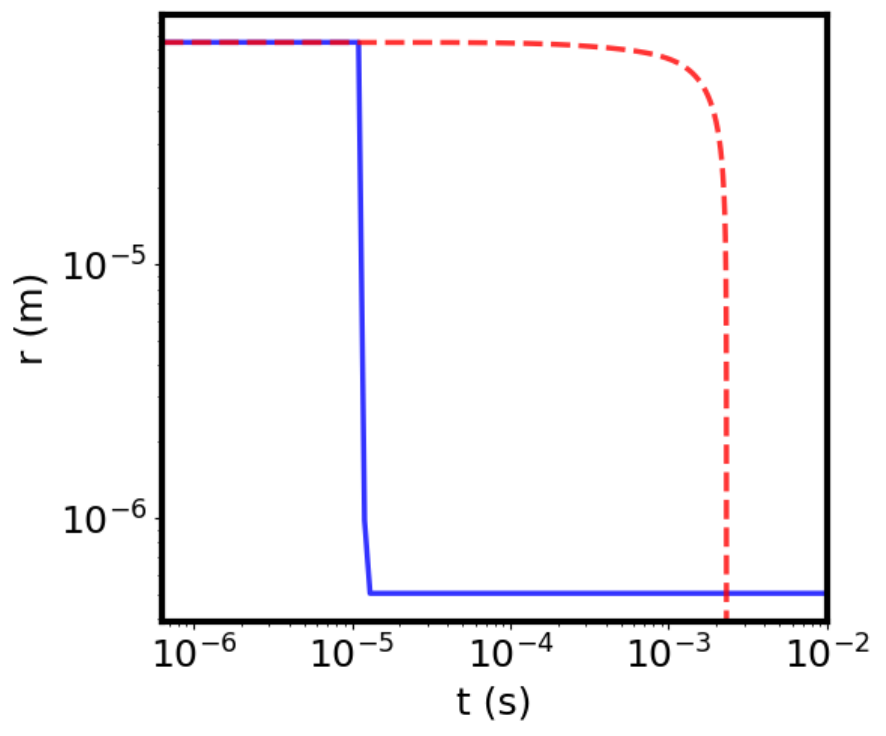}
    \caption{Evolution of droplet radius during spray processes: breakup only (evaporation disabled, {\color{black}\rule{8pt}{1pt}}) and evaporation only (breakup disabled, {\color{red}\rule{4pt}{1pt}\hspace{1pt}\rule{4pt}{1pt}\hspace{1pt}\rule{4pt}{1pt}}).}
    \label{fig:droplet-breakup}
\end{figure}

In the preceding sections describing the breakup and evaporation models, we assumed that breakup and evaporation can be decoupled -- an assumption which implies that these two phenomena occur at very disparate timescales. To examine the validity of this assumption, Fig.~\ref{fig:droplet-breakup} compares the temporal evolution of droplet diameter for breakup and evaporation processes, starting from an initial diameter of \SI{150}{\micro\meter}, with an initial droplet temperature of $300$~K, an ambient gas temperature of $350$~K, \textcolor{black}{with a droplet velocity of 112 m/s, obtained from the CFD simulation}. The breakup process (blue line) exhibits an abrupt reduction in droplet size due to the discrete nature of governing spray physics, with a rapid drop in droplet diameter from \SI{150}{\micro\meter} to approximately \SI{1}{\micro\meter} occurring at $t \approx 2\times10^{-5}$s. In contrast, the evaporation process (red dashed line) proceeds continuously, following the classical $D^2$-law \cite{godsave1953studies}, with near-complete droplet consumption at $t \approx 6\times10^{-3}$s.


The key observation is that breakup occurs on timescales much shorter than evaporation. For example, the time required for a droplet to shrink to one-tenth of its original size yields a ratio of $\tau_{\text{evaporation}} / \tau_{\text{breakup}} \approx 133$, demonstrating a separation exceeding 2 orders of magnitude. This confirms that for the problem considered in this study, droplet oscillation and distortion mechanisms act on far faster timescales than diffusion-controlled evaporation, thereby validating a sequential modeling approach in which breakup is first resolved, followed by evaporation applied to the post-breakup size distribution.

To further assess the robustness of this timescale separation across different thermal conditions, we repeated the same approach at an elevated ambient gas temperature of 900~K. At this higher temperature, the evaporation process accelerates significantly due to enhanced vapor pressure and increased diffusion rates. Despite the accelerated evaporation kinetics, the timescale separation remains substantial, with $\tau_{\text{evaporation}} / \tau_{\text{breakup}} \approx 21$ when evaluated at one-tenth of the initial droplet size. This reduction in the timescale ratio compared to the 350~K case reflects the stronger temperature dependence of evaporation relative to breakup, yet the separation still exceeds one order of magnitude, confirming that the decoupling assumption remains valid even under the elevated temperature conditions encountered in the combustor's high-temperature zones.

\subsection{CFD Flow Field Characteristics and Flame Stabilization}\label{sec:cfd_results}

\begin{figure*}[htbp]
\centering
\includegraphics[width=0.75\textwidth]{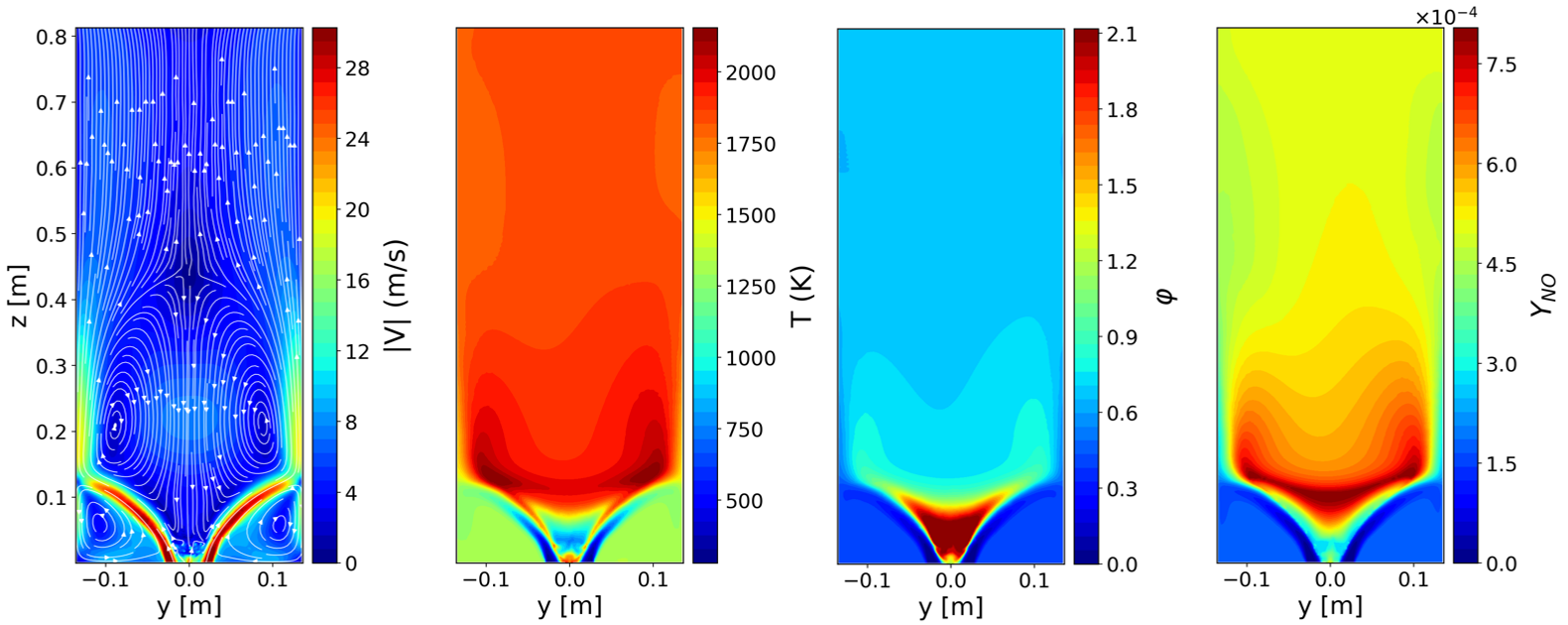}
\caption{Velocity flow field, Temperature, \textcolor{black}{Equivalence ratio}, and NO mass fraction contours obtained from the CFD simulation.}
\label{fig:temperature_contours}
\end{figure*}
\vspace{-2pt}
In Fig.~\ref{fig:temperature_contours}, slices of the flow field, temperature, \textcolor{black}{equivalence ratio, and} $Y_{\ce{NO}}$ are shown \textcolor{black}{based on an inlet temperature of 300K and the fuel mass flow rate of 2.2 g/s.} The flow field exhibits the formation of a pronounced central recirculation zone with reverse flow along the centerline, which provides the essential mechanism for flame stabilization through the recirculation of hot combustion products and active radicals \cite{syred1974}. A secondary recirculation zone develops \textcolor{black}{in the region surrounding the inflow}, establishing a hierarchical network of mixing regions that enhance fuel-oxidizer interaction. The temperature contour plot illustrates the thermal structure of the combustion process. The region directly above the nozzle is a fuel-rich region dominated by spray breakup and mixing, resulting \textcolor{black}{in relatively low temperatures}. This region is surrounded, in the radial direction, by a recirculation zone with moderately high temperatures, but still dominated by low temperature chemistry. This is evidenced by the relatively low concentrations of the OH radical, which is a marker for higher temperature heat release. Above the spray breakup region, a high temperature region with a characteristic V-shape that delineates the stoichiometric flame front location, can be observed. This zone is marked by high chemical reactivity, as evidenced by temperatures reaching maximum value of 2180 K and the OH radical reaching its peak concentrations. Beyond the axial locations of high reactivity, there is a gradual decline of temperature as the mixture is convected downstream, predominantly due to heat losses through the wall.

The NO mass fraction distribution shows distinct spatial patterns within the combustor. Peak NO concentrations occur within the high-temperature reaction zones, where temperatures exceed the threshold for significant thermal NO\textsubscript{x} formation (typically above 1800 K). The spatial distribution of NO shows elevated concentrations coincident with the primary flame zone rather than exhibiting substantial increases in the post-flame region, indicating that the residence time in high-temperature zones is the controlling factor for NO\textsubscript{x} formation under these operating conditions. The spatial distribution of NO mass fraction demonstrates the strong temperature dependence of the extended Zeldovich mechanism, where NO\textsubscript{x} formation requires both elevated temperatures and sufficient activation energy for nitrogen fixation reactions. The relatively confined nature of the high NO concentration regions suggests that the combustor design effectively limits the volume of high-temperature zones, thereby constraining overall NO\textsubscript{x} production. This spatial confinement of NO\textsubscript{x} formation to the immediate flame zone, rather than extensive formation in post-combustion regions, indicates favorable kinetic conditions for low-emission operation.

\subsection{Reactor Network Performance Validation}

\begin{figure}[H]
\centering
\includegraphics[width=0.4\textwidth]{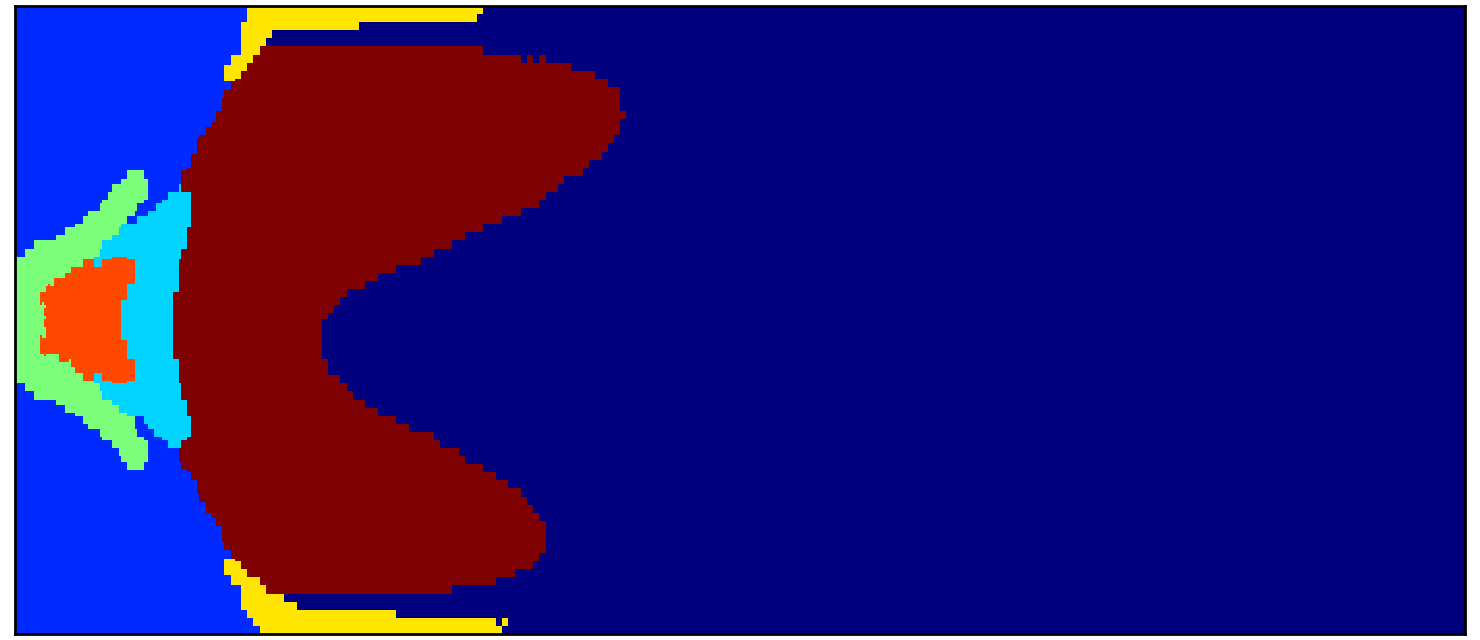}
\caption{$k$-means clustering partitioning of the combustor into seven zones: 
{evaporation and breakup (\color[HTML]{90EE90}\rule{5pt}{5pt}}), 
{mixing (\color[HTML]{FF8C00}\rule{5pt}{5pt}}), 
{flame 1 zone (PSR \color[HTML]{87CEEB}\rule{5pt}{5pt}}), 
{flame 2 zone (PSR \color[HTML]{DC143C}\rule{5pt}{5pt}}), 
{recirculation zone 1 (PSR \color[HTML]{0066FF}\rule{5pt}{5pt}}), 
{recirculation zone 2 (PSR \color[HTML]{FFD700}\rule{5pt}{5pt}}), and
{post-flame zone (PFR \color[HTML]{000080}\rule{5pt}{5pt}} ).} 
\label{fig:clustering}
\end{figure}

As discussed in Section \ref{sec:clustering_opt}, we employ a $k$-means clustering algorithm to partition the set of observations into clusters, which we subsequently map to their spatial locations to delineate the reactors. The spatial locations of the clusters, based on a 7-cluster case (representing the baseline), are shown in Fig.~\ref{fig:clustering}. \textcolor{black}{For this case, and for the other reactor counts considered later, the inlet reactor is termed the evaporator because it contains the fuel and air introduction region. The reactor containing the domain exit, which also has the largest axial extent, is treated as the plug-flow reactor. All remaining reactors are modeled as PSRs. In principle, each PSR can include droplet evaporation. Droplets enter the network, undergo breakup, heating, and evaporation, and are transported downstream until they fully evaporate, typically within or before the main reaction zone. Therefore,} although the partitioning is data-driven, the resulting clusters show some discernible patterns and can be labeled based on correlations to various flow and chemical regimes. The breakup/evaporation zone, for instance, is a V-shaped region directly above the nozzle, accounting for most of the spray breakup process. \textcolor{black}{This region corresponds to the V-shaped region of lowest temperature in the CFD contour plots of Fig. \ref{fig:temperature_contours}}. The mixing zone is a low-temperature \textcolor{black}{fuel-rich zone directly above the nozzle where the fuel} droplet continues to evaporate and mix with the surrounding hot gases, \textcolor{black}{and axially precedes a region of moderate reactivity, denoted as the flame 1 zone (with an average temperature of 1584 K)}. \textcolor{black}{The cluster with the highest reactivity (average T = 1977 K) exists directly above the flame 1 zone, coinciding with the regions of highest temperatures in Fig. \ref{fig:temperature_contours}. This is} followed in the axial direction by the post-flame region with gradually decreasing temperatures and combustion products. 
\vspace{-3pt}
\label{sec:reactor_performance}
\vspace{-2pt}
Finally, we also identify two clusters primarily as recirculation zones, referred to as recirculation zones 1 and 2, respectively. The recirculation and flame zones are modeled as perfectly stirred reactors, while the post-flame zone is approximated as a plug flow region, due to the dominance of the axial velocity component (see Fig.~\ref{fig:temperature_contours}). The spray breakup and mixing zones are modeled using customized reactors that account for the spray physics as described in Section \ref{sec:dev_rn}. 

\subsubsection{Sensitivity to Air Inlet Temperature}
\begin{figure*}[htbp]
\centering
\includegraphics[width=0.75\textwidth]{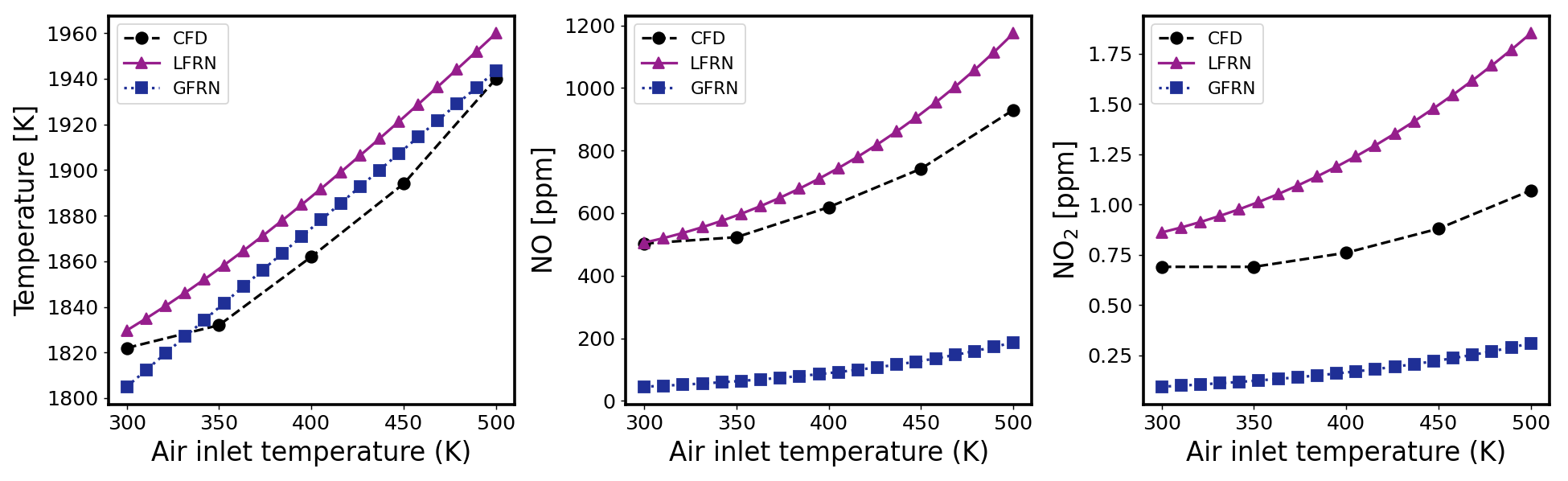}
\caption{Parametric analysis of the effect of inlet air temperature on exit (a) temperature, (b) \ce{NO} {concentration in ppm}, and (c) \ce{NO2} {concentration in ppm}.}
\label{fig:validation_comparison_temp}
\end{figure*}

Figure~\ref{fig:validation_comparison_temp} presents a comparison of reactor network modeling approaches against high-fidelity CFD simulations across a range of inlet air temperatures (300-500 K) \textcolor{black}{at a fuel flow rate of 2.2 g/s}. The comparative analysis evaluates three modeling approaches: the reference CFD simulation with detailed chemistry, the proposed liquid fuel reactor network (LFRN) incorporating spray dynamics, and a conventional gaseous fuel reactor network (GFRN) with the same number of reactors. \textcolor{black}{Different flamelet tables are applied for each temperature inlet condition, with each table generated by setting the air-side boundary condition in Z-coordinate to the air inlet CFD temperature}. The GFRN is developed in a similar fashion to the LFRN, with the same optimization procedure applied, but without the inclusion of spray dynamics (i.e., we assume that the fuel is introduced in the gaseous phase). We compare the capabilities of the proposed LFRN and the GFRN in accurately capturing the outlet temperature, \ce{NO}, and \ce{NO2} at the exit of the combustor.

For the results shown, we select the 300K case as the reference condition to tune the reactor as described in section \ref{sec:clustering_opt}. \textcolor{black}{Subsequent predictions at other inlet temperatures are computed based on the mass flow rates and residence times obtained from this reference condition without recalibration.} The results in Fig. \ref{fig:validation_comparison_temp} show that both reactor network methodologies (LFRN and GFRN) accurately capture the exit temperature, maintaining consistent agreement with CFD results throughout the investigated inlet temperature range. The thermal predictions exhibit deviations below \textcolor{black}{3\%} for both approaches, indicating that the global energy balance and heat release characteristics are adequately represented by both modeling frameworks. 
In contrast to the similar fidelities observed for the outlet temperatures, the {NO\textsubscript{x}} predictions reveal significant differences. For NO mass fraction predictions, the LFRN demonstrates good agreement with CFD results, \textcolor{black}{but errors grow from 0.54\% at the baseline to 26.3\% at 500 K}. \textcolor{black}{On the other hand,} the GFRN significantly underpredicts NO with errors of \textcolor{black}{80--91\%} throughout the entire temperature range.

This performance gap extends to NO\textsubscript{2} predictions. \textcolor{black}{Here, both methods struggle to quantitatively match the CFD results, but the LFRN performs better with} an accuracy within \textcolor{black}{46\%} of CFD values, while the GFRN demonstrates consistent underprediction ranging from \textcolor{black}{70\% to 80\%. In general,} the superior NO\textsubscript{x} prediction capability of the LFRN originates from its representation of spray-induced mixing heterogeneities and the resulting local variations in equivalence ratios. 
On the other hand, the GFRN's assumption of uniform mixing in the breakup-evaporation zones leads to lower global variations of equivalence ratio, resulting in a consistent underprediction of peak temperatures and lower NO\textsubscript{x} formation. This is discussed further in Section~\ref{sec:cluster_sensitivity}.

\subsubsection{Sensitivity to Fuel Flow Rate}  
\vspace{-2pt}
\begin{figure*}[htbp]
\centering
\includegraphics[width=0.75\textwidth]{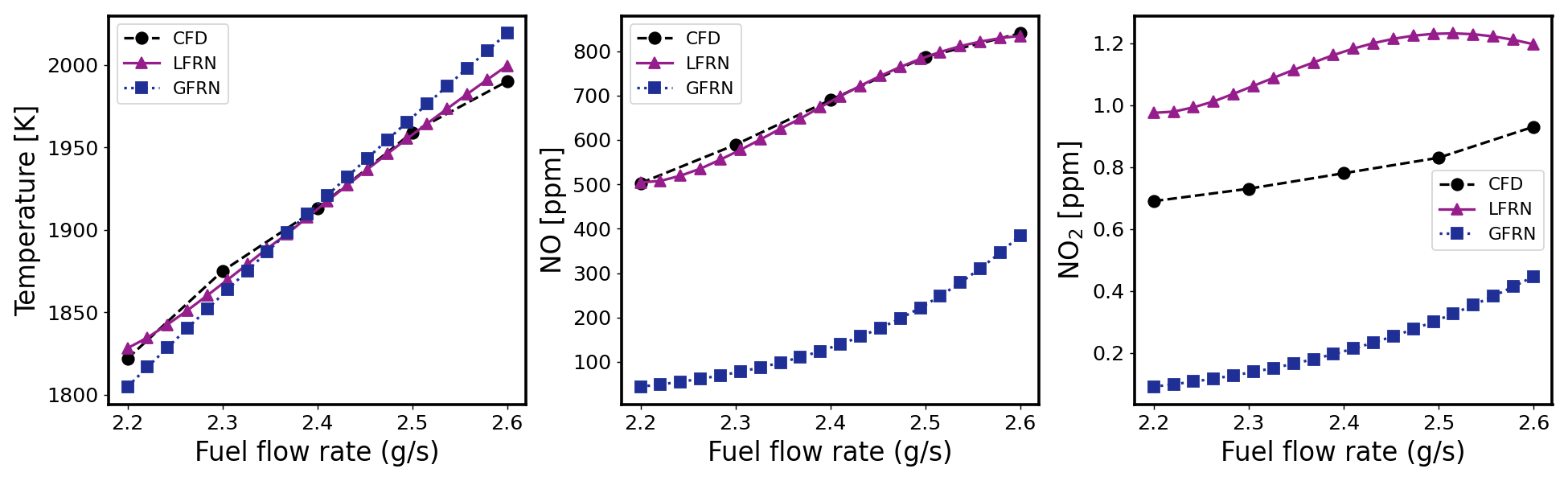}
    \caption{Parametric analysis of the effect of fuel flow rate on exit (a) temperature, (b) \ce{NO} {concentration in ppm}, and (c) \ce{NO2} {concentration in ppm}.}
    \label{fig:fuel_flow_comparison}
\end{figure*}  

The next parametric study examines the effect of varying the fuel flow rate on outlet temperature and \ce{{NO\textsubscript{x}}} predictions. In this case, changes in inlet mass flow rates modify the flow patterns in the reactor and the relative volumes of the reactors, meaning that a single calibration was insufficient. To address this, calibration was performed at the extreme mass flow rates of 2.2 g/s and 2.6 g/s. For all intermediate points, the $k$-means clustering results were interpolated to determine cluster centroids and volumes without additional optimization. Figure~\ref{fig:fuel_flow_comparison} illustrates the predictive capability of the reactor network across the 2.2–2.6~g/s operating range.  

Both reactor network approaches maintain excellent agreement with CFD results for thermal predictions, confirming that global heat-release characteristics are well captured independent of the spray-modeling methodology. However, similar to the inlet temperature parametric variation, significant discrepancies emerge in \ce{NO} predictions. The GFRN underestimates \ce{NO} mass fraction, with errors ranging from \textcolor{black}{54\%} under fuel-rich conditions to \textcolor{black}{91\%} at leaner conditions, while the LFRN shows a more accurate prediction of NO, with errors in the range of \textcolor{black}{0.29--0.67 \%}. While both methods generally capture the overall trend for \ce{NO2}, \textcolor{black}{the quantitative errors are higher}. However, the LFRN produces lower errors overall, ranging from \textcolor{black}{29 to 41\%}, while the GFRN underpredicts by \textcolor{black}{52 to 87\%}. {Extrapolation} beyond these bounds for the LFRN produced errors of 0.41\% and 4.2\% for outlet temperature and NO, respectively, at {2.0} g/s. At {2.7} g/s, the respective errors for these quantities were 1.2\% and 0.7\%.

\subsection{Sensitivity to the Number of Clusters}
\label{sec:cluster_sensitivity}

\subsubsection{Global Error Analysis}

\begin{figure}[H]
    \centering
    \includegraphics[width=0.4\textwidth]{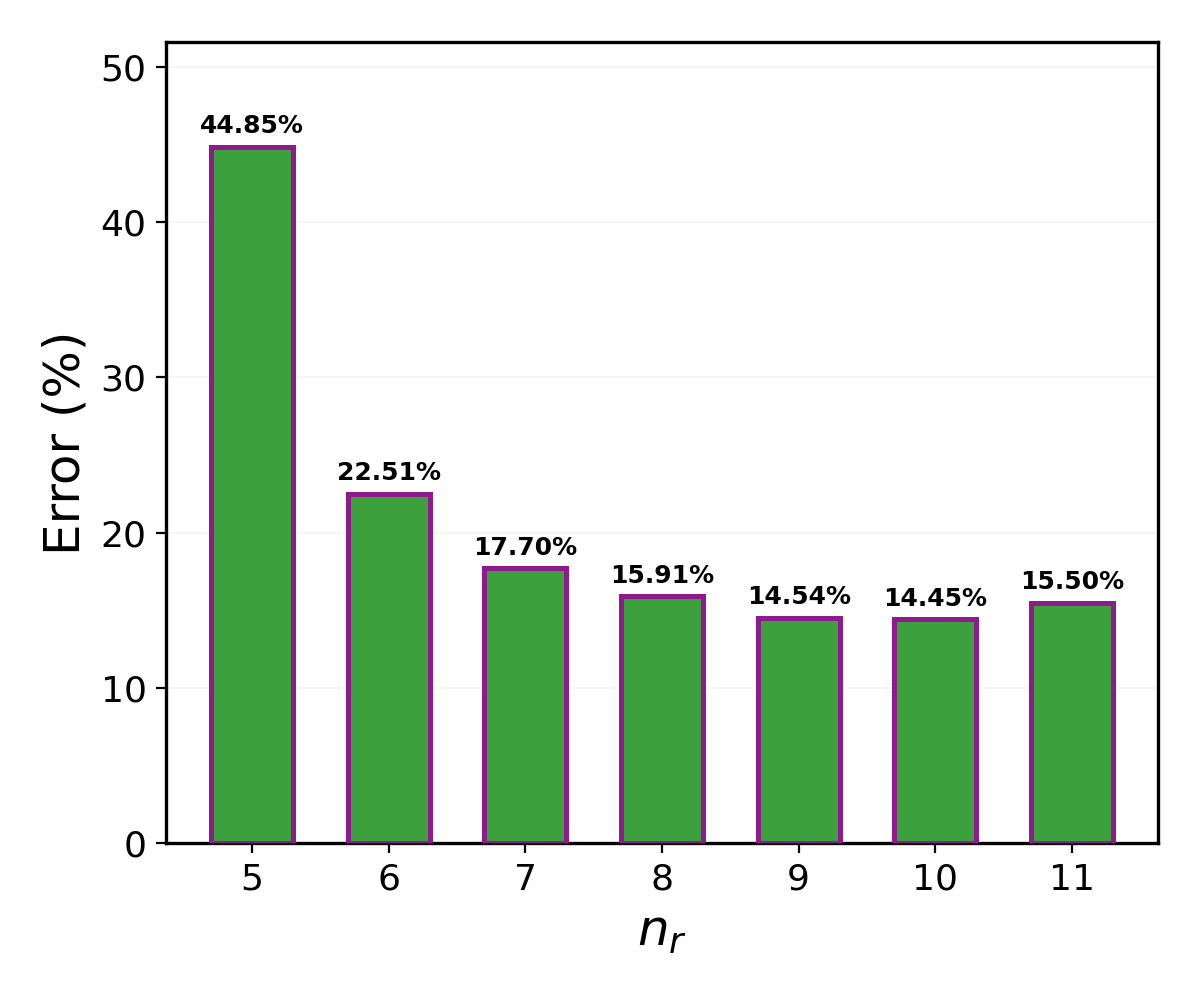}
    \vspace{-5pt}
    \caption{Normalized average {error, computed as the mean of the temperature and NO} errors for different reactor network configurations.}
    \label{fig:cluster_errors}
    
\end{figure}

One of the most important parameters in reactor network modeling is the number of reactors. In this section, we present a sensitivity study on how the accuracy of the LFRN approach varies with the number of clusters. The influence of network complexity on prediction accuracy was investigated by varying the number of clusters from 5 to \textcolor{black}{11}. Figure~\ref{fig:cluster_errors} quantifies the normalized average \textcolor{black}{error in NO and temperature, computed as $(\mathrm{error}_T+\mathrm{error}_{NO})/2$, as a function of the number of clusters, demonstrating convergence behavior with increasing spatial resolution.} To obtain representative errors across the entire CFD domain, the local errors were volume-weighted according to the size of each reactor zone, allowing larger reactors to influence the overall error in proportion to their volumes.  
\vspace{-8pt}
\vspace{-5pt}
\textcolor{black}{The results show a decrease in overall error as the number of clusters increases, followed by saturation beyond 9 clusters. This suggests that, beyond this point, the calibration procedure is unable to fully exploit the additional spatial resolution.} 

\subsubsection{{NO\textsubscript{x}} Formation Chemistry}

The superior {NO\textsubscript{x}} prediction capability of the LFRN compared to the GFRN can be understood through the thermal {NO\textsubscript{x}} formation mechanism. Under the high-temperature conditions in this combustor, {NO\textsubscript{x}} is produced primarily via the extended Zeldovich mechanism {~\cite{zeldovich1946oxidation}}:
{
\setlength{\abovedisplayskip}{6pt}
\setlength{\belowdisplayskip}{6pt}
\begin{align}
\ce{N2 + O &<=> N + NO} \label{eq:NO}
\end{align}
\vspace{-15pt}
\begin{align}
\ce{N + OH &<=> NO + H}
\end{align}
\vspace{-15pt}
\begin{align}
\ce{N + O2 &<=> NO + O} 
\end{align}
}
Eq. \textcolor{black}{\ref{eq:NO}} acts as a rate-limiting step due to its production of the N atom which is required to drive the remaining NO-producing reactions. Overall, this reaction exhibit{s} strong exponential temperature dependence ($E_a/R_u \approx 38,000\ K$), with NO formation rates rapidly increasing at higher temperatures. The LFRN's spray dynamics create localized fuel-rich zones that, upon ignition, generate peak temperatures exceeding \textcolor{black}{2200 K} with sufficient residence times for NO formation. In contrast, the GFRN's pre-mixed fuel assumption produces a homogeneous temperature field with relatively lower peak temperature (\textcolor{black}{$\approx$1800 K)}. This seemingly modest temperature reduction translates to a significant decrease in NO formation rates due to the exponential Arrhenius dependence, explaining the severe underpredictions in Figs.{~\ref{fig:validation_comparison_temp}} and{~\ref{fig:fuel_flow_comparison}}.

\subsubsection{Contour Plot Comparisons}
\vspace{-2pt}
\begin{figure*}[t]
\centering
\includegraphics[width=0.8\textwidth]{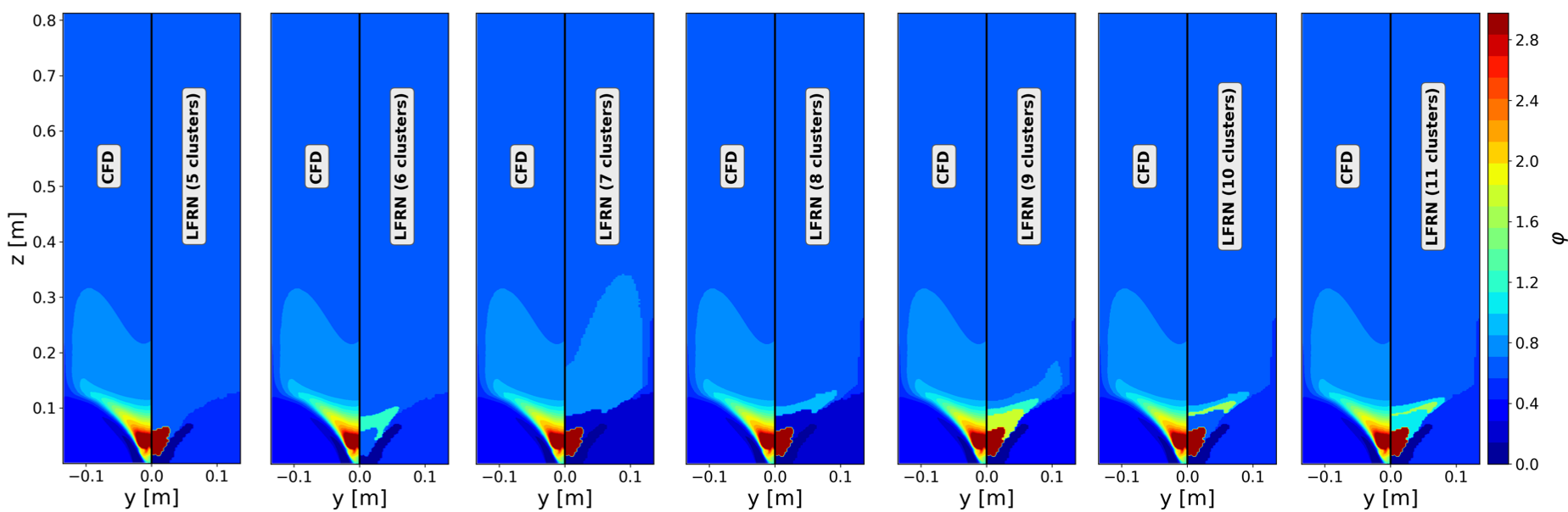}
    \caption{Comparison of LFRN cluster configurations with CFD equivalence ratio contours. Reactor network predictions (right half) versus CFD reference (left half) for 5–\textcolor{black}{11} clusters.}
    \label{fig:lfrn_equiv}
\end{figure*}

\begin{figure*}[t]
\centering
\includegraphics[width=0.8\textwidth]{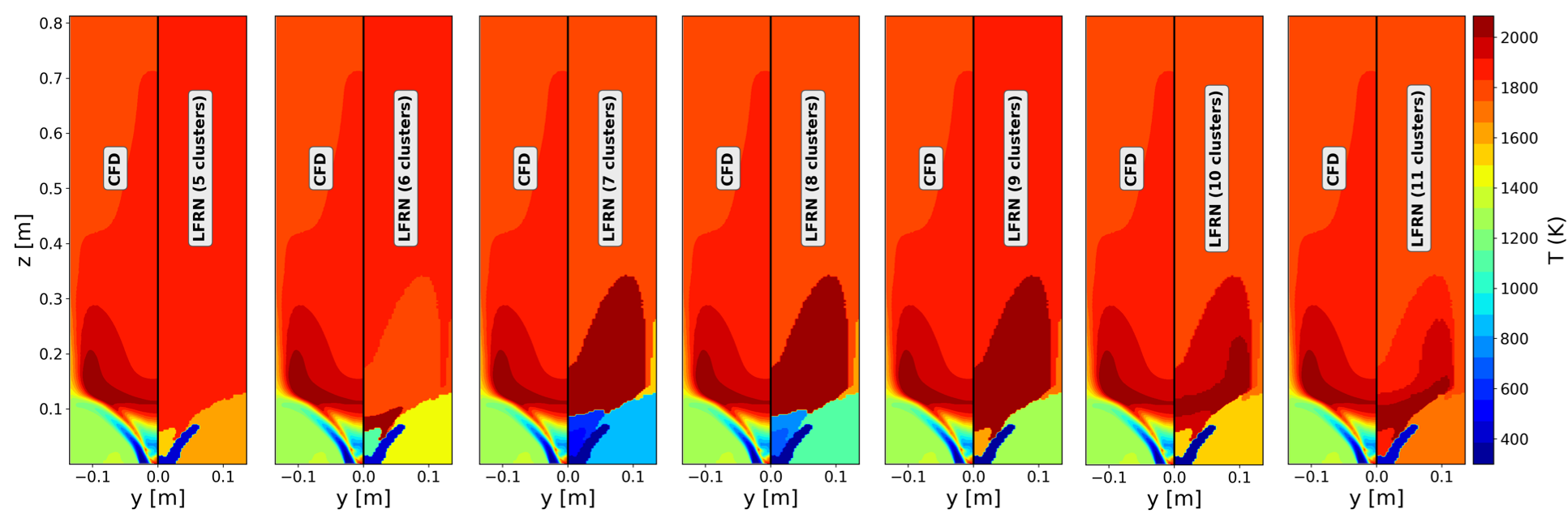}
    \caption{Comparison of LFRN cluster configurations with CFD temperature contours. Reactor network predictions (right half) versus CFD reference (left half) for 5–\textcolor{black}{11} clusters.}
    \label{fig:lfrn_temp}
\end{figure*}
\begin{figure*}[htbp]
\centering
\includegraphics[width=0.8\textwidth]{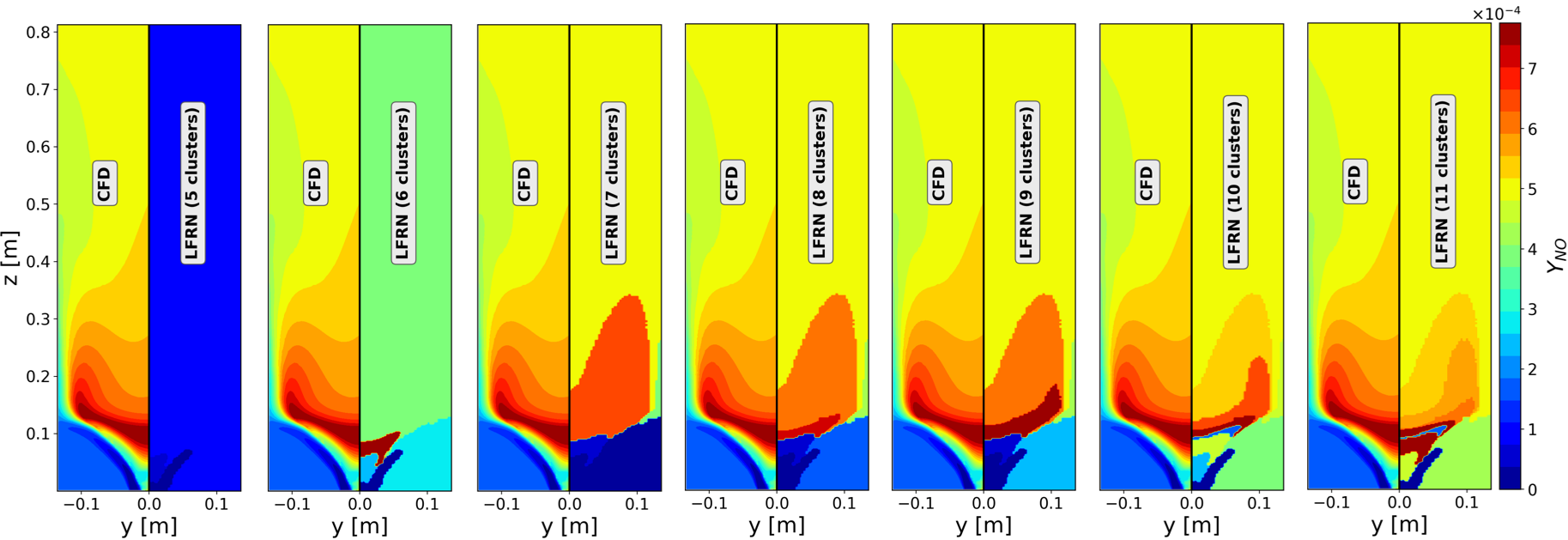}
    \caption{Comparison of LFRN cluster configurations with CFD \ce{NO} mass fraction contours. Reactor network predictions (right half) versus CFD reference (left half) for 5–\textcolor{black}{11} clusters.}
    \label{fig:lfrn_no}
\end{figure*}

Beyond the overall error metrics, we also examine the spatial fields by comparing contour plots of equivalence ratio, temperature, and \ce{NO} as predicted by the LFRN against CFD results (Figs.~\ref{fig:lfrn_equiv}, \ref{fig:lfrn_temp}, and \ref{fig:lfrn_no}). Each contour plot presents a split-domain visualization in which the right half shows the reactor network prediction and the left half displays the corresponding CFD reference solution, enabling direct qualitative comparison of spatial accuracy. \textcolor{black}{All the results shown are based on an air inlet temperature of 300K and a fuel mass flow rate of 2.2 g/s.} Overall, the plots demonstrate the progressive improvement in spatial field predictions as the number of clusters increases. 

The equivalence ratio distribution and \textcolor{black}{temperature plots (Fig.~\ref{fig:lfrn_equiv}, Fig.~\ref{fig:lfrn_temp})} illustrate the LFRN’s capability to capture spray-induced mixing heterogeneity and refine the reaction zones with increasing fidelity. The \textcolor{black}{5- and 6}-cluster configuration provide a reasonable match in the post-flame zone \textcolor{black}{but lack the resolution to capture the flame region.} Progressive refinement to 7-9 clusters significantly improves the representation \textcolor{black}{of the high temperature region, while also providing better resolution of the equivalence ratio in various regions, although the improvement is non-monotonic. Increasing to 10 and 11 reactors provides better resolution of the reaction zones, showing a hot core of peak reactivity surrounded by a slightly lower-temperature region of high reactivity.}


The \ce{NO} mass fraction distribution (Fig.~\ref{fig:lfrn_no}) provides the most stringent test of LFRN's fidelity, since pollutant formation depends sensitively on local temperature, residence time, and species concentrations. In the 5-cluster case, \ce{NO} mass fractions are significantly \textcolor{black}{underpredicted} in the flame region due to lower temperatures. \textcolor{black}{Refinement to 7-9 clusters provides improved resolution of the flame zone, leading to more accurate outlet \ce{NO} predictions, while further refinement to 10-11 clusters improves the representation of the regions extending from the evaporator to the flame zone.} \textcolor{black}{Results showing the contour plots for an intermediate mass flow rate of 2.4 g/s are shown in Figs. S1--S3 in the Supplementary Material.}


\begin{figure}[htbp]
\centering
\includegraphics[width=0.5\textwidth]{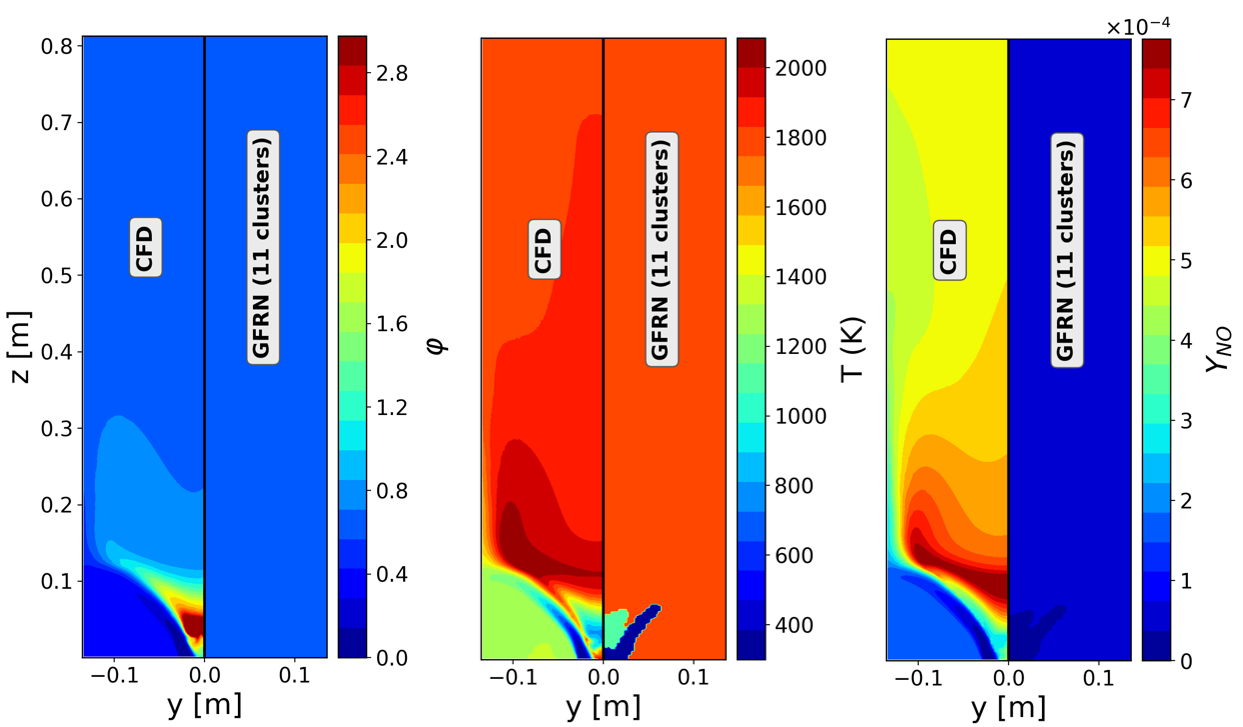}
\caption{Comparison of the \textcolor{black}{11}-cluster GFRN configuration with CFD-predicted equivalence ratio (left), temperature (middle), and \ce{NO} mass fraction.}
\label{fig:gfrn_temp}
\end{figure}

In contrast to the LFRN, the GFRN results, shown in Fig.~\ref{fig:gfrn_temp}, reveal the fundamental inadequacy of gaseous fuel reactor networks for the modeling of the considered liquid-fueled combustor. \textcolor{black}{Even when using a higher cluster count of 11,} analysis of the equivalence ratios of the GFRN (left) shows minimal spatial variation \textcolor{black}{from 0.652 to 0.653}. Compared to CFD, \textcolor{black}{the GFRN is unable to capture the fuel-rich region in the region directly above the nozzle and the subsequent gradual dilution as we approach the main reaction zone, } leading to an almost homogenized field. The CFD results show peak temperatures in a flame region that sits above the spray and mixing zones. In contrast, the GFRN \textcolor{black}{has lower peak temperatures that appear nearly constant in the reaction zones,} the recirculation zones and plug flow regions. This distortion of the temperature fields, in conjunction with the prescribed residence times, ultimately leads to quantitative differences in the \ce{NO} formation (right) within the combustor, such that the \ce{NO} mass fraction at the outlet is \textcolor{black}{more than an order of magnitude lower} than the corresponding CFD results \textcolor{black}{at the baseline condition}.

\subsection{Computational Efficiency Analysis}
\label{sec:computational_efficiency}
A comparison of runtimes for the CFD model and the reactor networks is presented in Table~\ref{tab:computational_comparison}. The results show that the LFRN can be executed in \textcolor{black}{$\mathcal{O}(10)$~s} on a single Intel Ice Lake (Intel\textsuperscript{\textregistered} Xeon\textsuperscript{\textregistered} Platinum 8358) CPU core, whereas the CFD simulation requires approximately \textcolor{black}{10~h on 256} cores of the same processor architecture. Even when run on far fewer processors, the 7-cluster LFRN achieves a speed-up factor \textcolor{black}{of about} \textcolor{black}{2,300$\times$} relative to the parallel CFD simulations. The most refined \textcolor{black}{11}-cluster LFRN configuration completes in only \textcolor{black}{35.8~s}, displaying a computational advantage that transforms combustor modeling from a resource-intensive activity to a practical design tool.

\begin{table}[H]
\centering
\caption{Computational performance comparison between CFD and reactor network approaches. Speed-up factors are reported relative to the CFD baseline of 10~h on 256 cores.}
\label{tab:computational_comparison}
\begin{tabular}{lccc}
\toprule
Method & \makecell{Computational\\Time} & Processors & \makecell{Speed-up\\Factor} \\
\midrule
CFD Simulation     & 10~h    & 256 & 1$\times$        \\
LFRN (5 clusters)  & 12.4~s  & 1   & 2{,}903$\times$  \\
LFRN (7 clusters)  & 15.6~s  & 1   & 2{,}308$\times$  \\
LFRN (11 clusters) & 35.8~s  & 1   & 1{,}006$\times$  \\
GFRN (5 clusters)  & 3.6~s   & 1   & 10{,}000$\times$ \\
GFRN (7 clusters)  & 5.5~s   & 1   & 6{,}545$\times$  \\
GFRN (11 clusters) & 9.5~s   & 1   & 3{,}789$\times$  \\
\bottomrule
\end{tabular}
\end{table}

\section{Conclusion}\label{sec:conclusions}
\vspace{-6pt}
In this study, a reactor network modeling framework for liquid-fueled combustors was introduced. The proposed approach incorporates simplified models for spray breakup, droplet heating, and evaporation into the conventional reactor network methodology. Validation was performed on a liquid-fueled can gas turbine combustor. Parametric studies were conducted by varying inlet air temperature and fuel flow rate. Analysis of the results showed that the proposed liquid-fueled reactor network (LFRN), compared to a standard gaseous-fueled reactor network (GFRN), provided substantially better agreement with CFD predictions for outlet temperature and \ce{{NO\textsubscript{x}}} emissions. A sensitivity study on the number of reactors demonstrated improved agreement with CFD as the number of clusters increased, \textcolor{black}{but with diminishing benefits beyond 9 clusters}. An analysis of the computational resources consumed showed a required time of \textcolor{black}{$\mathcal{O}(10)$~s} on a single CPU core, significantly lower than CFD compute requirements. As expected for a reduced-order model, the solution is valid only within a certain range of operating conditions. Nevertheless, this does not diminish its utility as a modeling tool, since its purpose is not to replace CFD in the design process, but to complement it. The LFRN can be calibrated against CFD results at selected operating points and then used to efficiently explore the proximal design space, reserving full-order CFD only for high-uncertainty regions. This strategy can enable shortened design cycles and more efficient use of computational resources.

\textcolor{black}{Although the proposed approach improves accuracy relative to conventional gaseous-fueled reactor networks, the current formulation has several limitations that provide directions for future development. The present formulation does not explicitly model droplet dynamics, including the coupled evolution of droplet position and velocity. Instead, spray locations are inferred from CFD data. In addition, the current implementation employs representative droplet sizes to characterize the spray. Future work will explore the incorporation of droplet size distributions, such as the Rosin-Rammler distribution \cite{rosinrammler1933}, as well as the development of more advanced spray models that account for droplet dynamics, including drag-induced transport and other relevant momentum-exchange processes. These enhancements are expected to further improve the predictive capability and generality of the proposed framework.}

\section*{Acknowledgments}
\vspace{-6pt}
This material is based upon work supported by the U.S. Department of Energy, Office of Energy Efficiency and Renewable Energy, Bioenergy Technologies Office under Award Number DE-SC0023463. The CONVERGE software licenses used in this work were provided through the CONVERGE Academic Program. The authors acknowledge the high-performance computing resources provided by Louisiana State University (\url{https://hpc.lsu.edu}) for enabling the computational work presented in this study.


\FloatBarrier

\bibliographystyle{unsrt}
\bibliography{references}

@book{pope2000turbulent,
  author    = {Pope, Stephen B.},
  title     = {Turbulent Flows},
  publisher = {Cambridge University Press},
  address   = {Cambridge, U.K.},
  year      = {2000}
}

@article{pitsch2006large,
  author  = {Pitsch, Heinz},
  title   = {Large-eddy simulation of turbulent combustion},
  journal = {Annu. Rev. Fluid Mech.},
  volume  = {38},
  pages   = {453--482},
  year    = {2006}
}

@article{fiorina2010modelling,
  author  = {Fiorina, B. and Vicquelin, R. and Auzillon, P. and Darabiha, N. and Gicquel, O. and Veynante, D.},
  title   = {A filtered tabulated chemistry model for {LES} of premixed combustion},
  journal = {Combust. Flame},
  volume  = {157},
  pages   = {465--475},
  year    = {2010}
}

@book{modest2013radiative,
  author    = {Modest, Michael F.},
  title     = {Radiative Heat Transfer},
  publisher = {Academic Press},
  year      = {2013}
}

@book{turns2011introduction,
  author    = {Turns, Stephen R.},
  title     = {An Introduction to Combustion: Concepts and Applications},
  publisher = {McGraw-Hill Education},
  year      = {2011}
}

@article{fichet2010reactor,
  author  = {Fichet, Vincent and Kanniche, Mohamed and Plion, Pierre and Gicquel, Olivier},
  title   = {A reactor network model for predicting {NO}x emissions in gas turbines},
  journal = {Fuel},
  volume  = {89},
  pages   = {2202--2210},
  year    = {2010}
}

@article{kaluri2018realtime,
  author  = {Kaluri, Abhishek and Malte, Philip C. and Novosselov, Igor V.},
  title   = {Real-time prediction of lean blowout using chemical reactor network},
  journal = {Fuel},
  volume  = {234},
  pages   = {797--808},
  year    = {2018}
}

@article{DeToni2013,
  author  = {De Toni Jr., Amir Roberto and Hayashi, Thamy and Schneider, Paulo Smith},
  title   = {A reactor network model for predicting {NO}x emissions in an industrial natural gas burner},
  journal = {J. Braz. Soc. Mech. Sci. Eng.},
  volume  = {35},
  pages   = {199--206},
  year    = {2013}
}

@article{Perpignan2019,
  author  = {Perpignan, Andr{\'e} A. V. and Sampat, Rishikesh and Gangoli Rao, Arvind},
  title   = {Modeling pollutant emissions of flameless combustion with a joint {CFD} and chemical reactor network approach},
  journal = {Front. Mech. Eng.},
  volume  = {5},
  pages   = {63},
  year    = {2019}
}

@article{Trespi2021,
  author  = {Trespi, Silvio and Nicolai, Hendrik and Debiagi, Paulo and Janicka, Johannes and Dreizler, Andreas and Hasse, Christian and Faravelli, Tiziano},
  title   = {Development and application of an efficient chemical reactor network model for oxy-fuel combustion},
  journal = {Energy Fuels},
  volume  = {35},
  pages   = {7121--7132},
  year    = {2021}
}

@article{Savarese2023,
  author  = {Savarese, Matteo and Cuoci, Alberto and De Paepe, Ward and Parente, Alessandro},
  title   = {Machine learning clustering algorithms for the automatic generation of chemical reactor networks from {CFD} simulations},
  journal = {Fuel},
  volume  = {343},
  pages   = {127945},
  year    = {2023}
}

@article{Savarese2024,
  author  = {Savarese, Matteo and Giuntini, Lorenzo and Malpica Galassi, Riccardo and Iavarone, Salvatore and Galletti, Chiara and De Paepe, Ward and Parente, Alessandro},
  title   = {Model-to-model {B}ayesian calibration of a chemical reactor network for pollutant emission predictions of an ammonia-fuelled multistage combustor},
  journal = {Int. J. Hydrog. Energy},
  volume  = {49},
  pages   = {586--601},
  year    = {2024}
}

@article{Villette2024,
  author  = {Villette, Sergios and Adam, Dimitris and Alexiou, Alexios and Aretakis, Nikolaos and Mathioudakis, Konstantinos},
  title   = {A simplified chemical reactor network approach for aeroengine combustion chamber modeling and preliminary design},
  journal = {Aerospace},
  volume  = {11},
  pages   = {22},
  year    = {2024}
}

@article{Dubal2024,
  author  = {D{\"u}bal, S{\"o}ren and Berkel, Leon L. and Debiagi, Paulo and Nicolai, Hendrik and Faravelli, Tiziano and Hasse, Christian and Hartl, Sandra},
  title   = {Chemical reactor network modeling in the context of solid fuel combustion under oxy-fuel atmospheres},
  journal = {Fuel},
  volume  = {364},
  pages   = {131096},
  year    = {2024}
}

@inproceedings{xu2013procedure,
  author    = {Xu, Kang and Shen, Suhua and Li, Chenkai and Zheng, Lipeng},
  title     = {A new procedure for predicting {NO}x emission in preliminary gas turbine combustor design},
  booktitle = {Proc. ASME Turbo Expo 2013: Turbine Technical Conf. Expo.},
  series    = {GT2013},
  pages     = {V01BT04A018},
  year      = {2013}
}

@misc{converge2025,
  author       = {Richards, K. J. and Senecal, P. K. and Pomraning, E.},
  title        = {{CONVERGE} 4.1},
  howpublished = {\url{https://convergecfd.com/}},
  note         = {Convergent Science, Madison, WI. Accessed March 12, 2026},
  year         = {2025}
}

@article{launder1974application,
  author  = {Launder, B. E. and Spalding, D. B.},
  title   = {The numerical computation of turbulent flows},
  journal = {Comput. Methods Appl. Mech. Eng.},
  volume  = {3},
  pages   = {269--289},
  year    = {1974}
}

@article{van2000flamelet,
  author  = {van Oijen, J. A. and de Goey, L. P. H.},
  title   = {Modelling of premixed laminar flames using flamelet-generated manifolds},
  journal = {Combust. Sci. Technol.},
  volume  = {161},
  pages   = {113--137},
  year    = {2000}
}

@book{peters2000turbulent,
  author    = {Peters, Norbert},
  title     = {Turbulent Combustion},
  publisher = {Cambridge University Press},
  address   = {Cambridge, U.K.},
  year      = {2000}
}

@article{fiorina2005modelling,
  author  = {Fiorina, B. and Gicquel, O. and Vervisch, L. and Carpentier, S. and Darabiha, N.},
  title   = {Approximating the chemical structure of partially premixed and diffusion counterflow flames using {FPI} flamelet tabulation},
  journal = {Combust. Flame},
  volume  = {140},
  pages   = {147--160},
  year    = {2005}
}

@inproceedings{orourke1987,
  author    = {O'Rourke, P. J. and Amsden, A. A.},
  title     = {The {TAB} method for numerical calculation of spray droplet breakup},
  booktitle = {{SAE} Tech. Pap. Ser.},
  pages     = {872089},
  year      = {1987}
}

@techreport{amsden1989,
  author      = {Amsden, A. A. and O'Rourke, P. J. and Butler, T. D.},
  title       = {{KIVA-II}: A computer program for chemically reactive flows with sprays},
  institution = {Los Alamos Natl. Lab.},
  number      = {LA-11560-MS},
  year        = {1989}
}

@article{rosinrammler1933,
  author  = {Rosin, P. and Rammler, E.},
  title   = {The laws governing the fineness of powdered coal},
  journal = {J. Inst. Fuel},
  volume  = {7},
  pages   = {29--36},
  year    = {1933}
}

@misc{cantera2022,
  author       = {Goodwin, David G. and Moffat, Harry K. and Schoegl, Ingmar and Speth, Raymond L. and Weber, Bryan W.},
  title        = {Cantera: An object-oriented software toolkit for chemical kinetics, thermodynamics, and transport processes},
  year         = {2022},
  note         = {Version 2.6.0, Zenodo}
}

@article{wang2018hychem1,
  author  = {Wang, Hai and Xu, Rui and Wang, Kai and Bowman, C. T. and Hanson, R. K. and Davidson, D. F. and Brezinsky, K. and Egolfopoulos, F. N.},
  title   = {A physics-based approach to modeling real-fuel combustion chemistry - {I}. Evidence from experiments, and thermodynamic, chemical kinetic and statistical considerations},
  journal = {Combust. Flame},
  volume  = {193},
  pages   = {502--519},
  year    = {2018}
}

@article{xu2018hychem2,
  author  = {Xu, Rui and Wang, Kai and Banerjee, Saurav and Shao, Ji and Parise, Tommaso and Zhu, Yiguang and Wang, Shaoheng and Movaghar, Arash and Lee, David J. and Zhao, Rui and Han, Xu and Gao, Yu and Lu, Tianfeng and Brezinsky, Kenneth and Egolfopoulos, Fokion N. and Davidson, Donald F. and Hanson, Ronald K. and Bowman, C. T. and Wang, Hai},
  title   = {A physics-based approach to modeling real-fuel combustion chemistry - {II}. Reaction kinetic models of jet and rocket fuels},
  journal = {Combust. Flame},
  volume  = {193},
  pages   = {520--537},
  year    = {2018}
}

@misc{wang2007uscmech,
  author       = {Wang, Hai and You, Xin and Joshi, A. V. and Davis, S. G. and Laskin, A. and Egolfopoulos, F. and Law, C. K.},
  title        = {{USC Mech Version {II}: High-Temperature Combustion Reaction Model of {H2/CO/C1-C4} Compounds}},
  howpublished = {\url{http://ignis.usc.edu/USC_Mech_II.htm}},
  note         = {Accessed March 12, 2026},
  year         = {2007}
}

@article{godsave1953studies,
  author  = {Godsave, G. A. E.},
  title   = {Studies of the combustion of drops in a fuel spray---the burning of single drops of fuel},
  journal = {Symp. (Int.) Combust.},
  volume  = {4},
  pages   = {818--830},
  year    = {1953}
}

@article{syred1974,
  author  = {Syred, N. and Beer, J. M.},
  title   = {Combustion in swirling flows: A review},
  journal = {Combust. Flame},
  volume  = {23},
  pages   = {143--201},
  year    = {1974}
}

@article{zeldovich1946oxidation,
  author  = {Zeldovich, Ya. B.},
  title   = {The oxidation of nitrogen in combustion and explosions},
  journal = {Acta Physicochim. URSS},
  volume  = {21},
  pages   = {577--628},
  year    = {1946}
}

@inproceedings{edwards2017reference,
  author    = {J.T. Edwards},
  title     = {Reference jet fuels for combustion testing},
  booktitle = {55th AIAA Aerospace Sciences Meeting},
  address   = {Grapevine, TX},
  year      = {2017}
}

@article{du2014equivalent,
  title={Equivalent reactor network model for simulating the air gasification of polyethylene in a conical spouted bed gasifier},
  author={Du, Yupeng and Yang, Qi and Berrouk, Abdallah S and Yang, Chaohe and Al Shoaibi, Ahmed S},
  journal={Energy fuels},
  volume={28},
  pages={6830--6840},
  year={2014},
  publisher={ACS Publications}
}

@book{sirignano2010fluid,
  author    = {W.A. Sirignano},
  title     = {Fluid Dynamics and Transport of Droplets and Sprays},
  publisher = {Cambridge University Press},
  address   = {Cambridge},
  year      = {2010}
}

@book{sazhin2014droplets,
  author    = {S.S. Sazhin},
  title     = {Droplets and Sprays},
  publisher = {Springer},
  address   = {London},
  year      = {2014}
}

@article{hansen2001cmaes,
  author  = {N. Hansen and A. Ostermeier},
  title   = {Completely derandomized self-adaptation in evolution strategies},
  journal = {Evol. Comput.},
  volume  = {9},
  pages   = {159--195},
  year    = {2001},
}

\end{document}